\documentclass[aps,12pt]{revtex4}

\begin{document}

\tighten


\title{Chern-Simons Spinor Electrodynamics in the Light-Cone Gauge}

\author{Wenfeng Chen}
\email{wenfengc@nipissingu.ca}
 \affiliation{Department of Mathematics, Nipissing University, North Bay,
Ontario, Canada P1B 8L7}

\begin{abstract}
\noindent The one-loop quantum corrections of Chern-Simons spinor
electrodynamics in the light-cone gauge has been investigated. We
have calculated the vacuum polarization tensor, fermionic
self-energy and on-shell vertex correction with a hybrid
regularization consisting of a higher covariant derivative
regularization and dimensional continuation. The
Mandelstam-Leibbrandt prescription is used to handle the spurious
light-cone singularity in the gauge field propagator. We then
perform the finite renormalization to define the quantum theory. The
generation of the parity-even Maxwell term and the arising  of
anomalous magnetic moment from quantum corrections are reproduced as
in the case of a covariant gauge choice. The Ward identities in the
light-cone gauge are verified to satisfy explicitly. Further, we
have found the light-cone vector dependent sector of local quantum
effective action for the fermion is explicitly gauge invariant, and
hence the covariance of $S$-matrix elements of the theory can be
achieved.


\end{abstract}

\maketitle

\vspace{3ex}


\section{Introduction}

The first step in calculating quantum correction of a gauge theory
by perturbation theory is choosing a gauge condition to eliminate
the non-physical degrees of freedom caused by gauge symmetry. This
process is called gauge fixing. Despite that physically measurable
results should be independent of gauge choice, but with different
gauge-fixing, the quantum theory presents  distinct features in both
calculation  techniques and the resultant quantum corrections. The
usually preferred choice is a Lorentz covariant gauge condition like
$\partial_\mu A^\mu=0$, since the Lorentz covariance can be
preserved in the entire calculation process, and further, the
propagator of gauge field has a nice analytical structure.

Nevertheless, in certain circumstances, a non-covariant gauge choice
turns out to be more convenient than a covariant one, since this
kind of gauge choice can somehow approach to physical degrees of
freedom straightforwardly at classical level. Especially, a
non-covariant gauge fixing in a non-Abelian gauge theory can make
ghost fields decouple from physical sector in classical stage, and
avoid the notorious Gribov's ambiguity haunted the gauge-fixing
procedure \cite{banaso, glbook}.

However, a non-covariant gauge fixing brings about a spurious
singularity in the gauge field propagator \cite{banaso, glbook}.
This hinders the loop integration in perturbation theory from being
performed straightforwardly as in the covariant case. Therefore, a
prescription of handling the spurious singularity must be defined so
that the denominator of the integrand in a loop integration is
quadratic  in the loop momentum \cite{banaso, glbook}. A number of
prescriptions had been proposed \cite{banaso, glbook}. Up to now it
seems that the most convenient and universal  prescription is the
$n_\mu^\ast$-prescription suggested by Mandelstam \cite{mand} and
Leibbrandt \cite{george}, which is now  termed as the
Mandelstam-Leibbrandt (ML) prescription \cite{banaso}. It has been
tested that the ML prescription can give consistent results for any
non-covariant gauge choices at one-loop level for gauge theories in
both four and three dimensions \cite{glbook}, although its
applicability in evaluating two-loop and higher order quantum
corrections needs to be verified explicitly.

The study on the pure non-Abelian Chern-Simons (CS) gauge theory in
the light-cone gauge at one-loop  with the ML-prescription was
pioneered by Martin and Leibbrandt \cite{martin2}. A consistent
result with the covariant gauge fixing had been achieved: the
celebrated finite quantum correction $k$-shift \cite{kshift,
martin1,glwchen} of the gauge coupling is reproduced, and the
non-local gauge dependent terms are unobservable. Consequently, the
topological feature of the theory is preserved. Hence the
applicability  of the ML-prescription to three-dimensional gauge
theory with parity violation had been verified at one-loop order
\cite{martin2}.

In this article we shall investigate three-dimensional Chern-Simons
spinor electrodynamics \cite{kogan, ccl} in the light-cone gauge,
i.e., $U(1)$ CS gauge theory coupled with fermions. This model has
some distinct features from the pure non-Abelian CS gauge theory,
and it is worthy to observe its quantum corrections in the
light-cone gauge with the ML prescription. First, it is not a
topological field theory since the coupling of gauge field with
fermion requires an explicit involvement of space-time metric, and
the theory has local dynamical degrees of freedom. Second, from the
one-loop result of four-dimensional gauge theory in the light-cone
gauge calculated with the ML-prescription, the light-cone vector
dependent part in the local quantum effective action for fermions
should take a specific gauge-invariant form \cite{banaso},
determined by the Ward-Takahashi identities in the light-cone gauge,
so that the covariance of $S$-matrix elements of theory can be
recovered. It is interesting to check explicitly whether such a
result arises in a three-dimensional gauge theory in the light-cone
gauge. Third, in contrast to the pure non-Abelian CS gauge theory,
which has only one dimensionless parameter -- the gauge coupling, CS
spinor electrodynamics has a parameter with  mass dimension -- the
mass of the fermion. This will make both the tensor structure and
form factors of quantum corrections of the theory much more
involved.

Furthermore, it has been shown that in the covariant gauge-fixing,
 CS spinor electrodynamics presents some remarkable
 quantum effects including the generation of Maxwell (or parity-even) term
 and  the arising of anomalous magnetic moment of the fermion
 \cite{kogan}. It is significant to observe these radiative
 corrections in the light-cone gauge with the ML prescription, since
 this can not only  reveal quantum features of Chern-Simons-matter theory,
 but also confirm and consolidate the validity of the ML
 prescription in evaluating quantum corrections of three-dimensional
 gauge theories in the light-cone gauge.

In Sect.\,\ref{sectmodel}, we introduce the classical CS spinor
electrodynamics with the light-cone gauge-fixing. For later
perturbative calculation, we choose a hybrid regularization scheme
to derive the Feynmann rules. The hybrid regularization is a
combination of higher covariant derivative regularization and
dimensional continuation with the Maxwell term as the higher
derivative term. Sect.\,\ref{twopf} contains a calculation on
two-point functions at one-loop including the vacuum polarization
tensor $\Pi_{\mu\nu}(p)$ and the fermionic self-energy $\Sigma (p)$.
We use the ML prescription to handle the spurious singularity of the
gauge field propagator. In Sect.\,\ref{osvertex} we display a
detailed evaluation of one-loop quantum vertex on the mass-shell of
the fermion. Because it requires two light-cone vectors $n_\mu$ and
$n_\mu^\ast$ to implement the ML prescription,  the calculation on
the form factors of on-shell vertex correction is much more tedious
than the case of covariant gauge-fixing. In Sect.\,\ref{renorma} we
perform renormalization on the quantum corrections found in previous
two sections, and reveal quantum effects and the structure of local
quantum effective action of the theory. The calculation techniques
and integration formulas are given in details in Appendices
\ref{appa} and \ref{appb}. In Appendix \ref{appc} we derive the Ward
identities of CS spinor electrodynamics in the light-cone gauge. In
particular, we show the explicit restriction of the Ward identity on
the general form of two-point function of gauge field, and  the
relation between the gauge field-fermion-fermion vertex correction
and the fermionic self-energy.

\section{Chern-Simons Spinor Electrodynamics in the Light-Cone Gauge}
\label{sectmodel}

The Lagrangian density of CS spinor electrodynamics in the
light-cone gauge is
\begin{eqnarray}
{\cal L}=\frac{1}{2}\epsilon^{\mu\nu\rho}A_\mu\partial_\nu
A_\rho+\overline{\psi}(i \partial\hspace{-2.4mm}/+e A
\hspace{-2.4mm}/-m)\psi-\frac{1}{2\xi}\left(n^\mu A_\mu \right)^2,
\label{sedeq1}
\end{eqnarray}
where $(n_\mu) = (n_0,n_1,n_2)$ is the light-cone vector, which by
defination satisfies $n^2=0$, and further $\xi\rightarrow 0$. The
$\gamma$-matrices in the Lagrangian density (\ref{sedeq1}) are
chosen as follows:
\begin{eqnarray}
\gamma^0=\sigma_2, ~~\gamma^1=i\sigma_3, ~~\gamma^2=i\sigma_1.
\end{eqnarray}
Consequently, the algebra formed by the $\gamma$-matrices is
\begin{eqnarray}
&&
\gamma_\mu\gamma_\nu=g_{\mu\nu}-i\epsilon_{\mu\nu\rho}\gamma^\rho,
~~~\left\{\gamma_\mu,\gamma_\nu  \right\}=2g_{\mu\nu},\nonumber\\
&& \left[\gamma_\mu,\gamma_\nu
\right]=-2i\epsilon_{\mu\nu\rho}\gamma^\rho,~~~
\left(g_{\mu\nu}\right)=\mbox{diag}(1,-1,-1).
\label{gamma}
\end{eqnarray}
The gauge-fixing term $-1/(2\xi)\,(n^\mu A_\mu)^2$ in the Lagrangian
density (\ref{sedeq1}) comes from the light-cone gauge condition
$n^\mu A_\mu=0$, $n^2=0$
with $\xi\rightarrow 0$ in the gauge field propagator.

To investigate the perturbative quantum corrections of CS spinor
dynamics, we must choose a regularization scheme to deal with the
ultraviolet divergence in loop integration. Usually the most
convenient method is dimensional regularization. However, due to the
particular feature of CS term: its kinetic operator
$\epsilon^{\mu\nu\rho}\partial_\rho$ being a first-order
non-positive definite differential operator, we must first implement
a higher covariant derivative regularization scheme. The simplest
gauge invariant higher covariant derivative term is the Maxwell
term,
\begin{eqnarray}
{\cal L}_{\Lambda}=-\frac{1}{4\Lambda} F_{\mu\nu}F^{\mu\nu},
\end{eqnarray}
where $F_{\mu\nu}=\partial_{\mu}A_\nu-\partial_{\nu}A_\mu$ and
$\Lambda$ is the regulator.

To apply dimensional regularization, we should use the 't
Hooft-Veltman prescription to define the dimensional continuation of
$\epsilon_{\mu\nu\rho}$ tensor and the $\gamma$-matrices
\cite{hooft, breit}.The regularized $d$-dimensional space is divided
into a  direct sum of the original three-dimensional space and a
($d-3$)-dimensional space, $d$ being a complex number \cite{martin1,
martin2}. However, for the Abelian CS theory,  the $\epsilon$-tensor
appears only in the gauge field propagator, and especially, in this
work we consider only the perturbative theory at one-loop level.
Hence the 't Hooft-Veltman recipe makes no difference with the usual
na\"{i}ve dimensional continuation, and the inconsistency found in
Ref.\,\cite{chen1} will not arise. The explicit calculations carried
out later will confirms this argument.

As a hybrid combination of the dimensional continuation and the
higher covariant derivative regularization, the order of removing
the regulators after the renormalization is first taking the limit
$d\rightarrow 3$ and then $\Lambda\rightarrow \infty$.

The regularized Lagrangian density ${\cal L}+{\cal L}_\Lambda$ leads
to the following  tree-level Feynman rules:
\begin{itemize}
\item Photon propagator \cite{martin2}:
\begin{eqnarray}
iG^{(0)}_{\mu\nu}(p,n,\Lambda)&=& \frac{i\Lambda}{p^2
(p^2-\Lambda^2)} \left[i\Lambda \epsilon_{\mu\nu\rho}p^\rho
-\frac{i\Lambda}{n\cdot p} \left(p_\mu
\epsilon_{\nu\alpha\beta}-p_\nu \epsilon_{\mu\alpha\beta} \right)
p^\alpha n^\beta\right.\nonumber\\
&&\left.-p^2 g_{\mu\nu}+\frac{p^2}{n\cdot p}(p_\mu n_\nu+p_\nu n_\mu) \right]
\nonumber\\
&=& \frac{i\Lambda}{(p^2-\Lambda^2)n\cdot p}\left[
i\Lambda\epsilon_{\mu\nu\rho}n^\rho- n\cdot p g_{\mu\nu}+
\left(p_\mu n_\nu+p_\nu n_\mu \right)\right],
\label{phopro}
\end{eqnarray}
where the following one of Martin's identities has been used
\cite{martin2,martin3},
\begin{eqnarray}
\frac{1}{n\cdot p} \epsilon_{\mu\nu\rho}n^\rho =\frac{1}{p^2}
\epsilon_{\mu\nu\rho}p^\rho-\frac{1}{p^2 (n\cdot p)}
\left(p_\mu \epsilon_{\nu\alpha\beta}-p_\nu \epsilon_{\mu\alpha\beta} \right)p^\alpha n^\beta.
\label{mid1}
\end{eqnarray}
As $\Lambda\longrightarrow\infty$ at tree-level, the propagator (\ref{phopro}) reduces to
 \begin{eqnarray}
 iG^{(0)}_{\mu\nu}(p,n)&=&\frac{1}{n\cdot p}\epsilon_{\mu\nu\rho}{n^\rho}.
 \end{eqnarray}
\item Fermionic propagator:
\begin{eqnarray}
iS^{(0)}(p)=i\frac{p\hspace{-1.7mm}/+m}{p^2-m^2}\,.
\end{eqnarray}
\item Gauge field-fermion-fermion vertex:
\begin{eqnarray}
-ie\Gamma_\mu^{(0)}(p,q,r)=-ie\gamma_\mu (2\pi)^d
\delta^{(d)}(p+q+r).
\end{eqnarray}

\end{itemize}

In the following sections we shall calculate one-loop quantum
corrections of the theory and  show quantum features of CS spinor
electrodynamics in the light-cone gauge.

\section{One-loop Vacuum Polarization Tensor and  Fermionic Self-energy}
\label{twopf}

\subsection{Vacuum Polarization Tensor}

Since the characteristic of the light-cone gauge fixing involves
only in $U(1)$ CS gauge field propagator, the vacuum polarization
tensor is identical to that in the usual covariant gauge,
\begin{eqnarray}
&& i\Pi^{(1)}_{\mu\nu}(p^2)= -e^2\int\frac{d^dk}{(2\pi)^d}
\frac{\mbox{Tr}\left[ \gamma_\nu (k\hspace{-2mm}/ +
p\hspace{-1.8mm}/ + m)
\gamma_\mu (k\hspace{-2.1mm}/ + m) \right]}{(k^2-m^2)[(k+p)^2-m^2]}\nonumber\\
&=& -2e^2\int\frac{d^dk}{(2\pi)^d}\frac{-im \epsilon_{\mu\nu\rho}p^\rho
+2 k_\mu k_\nu+k_\mu p_\nu+k_\nu p_\mu-g_{\mu\nu}
\left[k\cdot (k+p)-m^2 \right]}{(k^2-m^2)[(k+p)^2-m^2]},
\end{eqnarray}
where we have used  $\gamma$-matrix algebra listed in (\ref{gamma}).
The parity-odd part is finite, and one can take the limit
$d\rightarrow 3$ before performing the loop integration. The
parity-even part contains the superficially linear and logarithmic
divergent terms, which can be evaluated by the dimensional
regularization. Using the formula listed in Appendix \ref{appb}, we
obtain (after taking the limit $d\to 3$)
\begin{eqnarray}
\Pi^{(1)}_{\mu\nu}(p)&=& i\epsilon_{\mu\nu\rho}p^\rho\Pi_{\rm
o}(p^2)+
\left(p^2g_{\mu\nu}-p_\mu p_\nu\right)\Pi_{\rm e}(p^2)\nonumber\\
&=& \frac{e^2}{4\pi}\left\{i\epsilon_{\mu\nu\rho}p^\rho
\frac{m}{p}\ln \left[\frac{1+p/(2m)}{1-p/(2m)}\right]\right.\nonumber\\
&-&\left. \left(p^2g_{\mu\nu}-p_\mu p_\nu\right) \frac{1}{m}\left[
-\frac{m^2}{p^2}+\frac{m}{p}\left(\frac{1}{4}+ \frac{m^2}{p^2}\right)
\ln\left(\frac{1+p/(2m)}{1-p/(2m)}\right)
 \right]\right\}.
 \label{vpt1}
\end{eqnarray}
In Eq.\,(\ref{vpt1}) $p\equiv |p|$,  and $\Pi_{\rm o}(p^2)$ and
$\Pi_{\rm e}(p^2)$ represent the parity odd- and even form factors
of the vacuum polarization tensor,
\begin{eqnarray}
\Pi_{\rm o}(p) &=&  \frac{e^2}{4\pi}\frac{m}{p}
\ln \left[\frac{1+p/(2m)}{1-p/(2m)}\right]\, ,\nonumber\\
\Pi_{\rm e} (p) &=&  \frac{e^2}{4\pi}\frac{1}{m}\left[
\frac{m^2}{p^2}-\frac{m}{p}\left(\frac{1}{4}+ \frac{m^2}{p^2}\right)
\ln\left(\frac{1+p/(2m)}{1-p/(2m)}\right)\right].
\label{vpt2}
\end{eqnarray}

\subsection{Self-energy of Fermion}

Compared with the case of the covariant Landau gauge \cite{kogan},
the fermionic self-energy has some distinct features due to the
presence of the spurious light-cone gauge singularity $1/(n\cdot k)$
in the propagator of $U(1)$ CS gauge field:
\begin{eqnarray}
-i\Sigma^{(1)} (p,m,n,\Lambda,d)&= & e^2\int
\frac{d^dk}{(2\pi)^d}\left\{ \frac{\gamma^\nu (k\hspace{-2mm}/ +
p\hspace{-1.8mm}/ + m)\gamma^\mu \Lambda}{[(k+p)^2-m^2]
(k^2-\Lambda^2) n\cdot k}
\right.\nonumber\\
&&\times \left.\left[
i\Lambda\epsilon_{\mu\nu\rho}n^\rho- n\cdot k g_{\mu\nu}+
\left(k_\mu n_\nu+k_\nu n_\mu \right)\right]\right\}\nonumber\\
&=& e^2\int \frac{d^dk}{(2\pi)^d} \left\{-\frac{\Lambda \gamma^\mu
(k\hspace{-2mm}/ + p\hspace{-1.8mm}/ + m)\gamma_\mu}{[(k+p)^2-m^2]
(k^2-\Lambda^2)} \right.\nonumber\\
&&+\frac{i\Lambda^2\epsilon^{\mu\nu\rho}n_\rho \gamma_\nu
(k\hspace{-2mm}/ + p\hspace{-1.8mm}/ + m)\gamma_\mu }
{[(k+p)^2-m^2] (k^2-\Lambda^2) n\cdot k}\nonumber\\
&&\left.+\frac{ \Lambda n\hspace{-2mm}/ (k\hspace{-2mm}/ + p\hspace{-1.8mm}/ + m)
k\hspace{-2mm}/ + k\hspace{-2mm}/ (k\hspace{-2mm}/ + p\hspace{-1.8mm}/ + m)
 n\hspace{-2mm}/ }{[(k+p)^2-m^2] (k^2-\Lambda^2) n\cdot k}\right\}.
\end{eqnarray}

 The spurious light-cone gauge singularity $1/(n\cdot
k)$ in the integrand brings difficulty in evaluating the loop
integration. We use the ML prescription in Minkowskian space to
handle the singularity \cite{mand, george}:
\begin{eqnarray}
\frac{1}{n\cdot k}&=& \lim_{\epsilon\to 0}\frac{n^\ast\cdot k}
{(n\cdot k)(n^\ast\cdot k)+i\epsilon}=\lim_{\epsilon\to 0}
\frac{n^\ast\cdot k}{n_0^2k_0^2+({\bf n}\cdot {\bf k})^2+i\epsilon}, ~~~\epsilon >0, \nonumber\\
n&=& \left( n_\mu\right)=(n_0,{\bf n}), ~~~n^\ast= (n_\mu^\ast)=
(n_0, -{\bf n}), ~~~ n_0 >0. \label{mlpres}
\end{eqnarray}
Obviously, $n^\ast$ is also a light-cone vector since $n^{\ast
2}=0$.

 We first expand the numerator of each term in the integrand
using the $\gamma$-matrix algebra (\ref{gamma}), and then separate
the integrands into the parts with and without the light-cone pole,
\begin{eqnarray}
-i\Sigma^{(1)} (p,m, n, \Lambda, d)&=& -i \left(\Sigma_{\rm
NP}+\Sigma_{\rm P} \right)\, , \nonumber\\
-i \Sigma_{\rm NP} &=& e^2\int \frac{d^dk}{(2\pi)^d}\frac{2\Lambda^2
+\Lambda \left[(d-2) k\hspace{-2mm}/ - (4-d)  p\hspace{-1.8mm}/
-(d-2) m
\right]}{[(k+p)^2-m^2] (k^2-\Lambda^2)}\, ,\nonumber\\
-i \Sigma_{\rm P} &=& e^2\int \frac{d^dk}{(2\pi)^d} \frac{2\Lambda^2
(n\cdot p-m n\hspace{-2mm}/ ) + 2 \Lambda \left[ k\cdot (k+p)
n\hspace{-2mm}/ +n\cdot p k\hspace{-2mm}/ \right]}{[(k+p)^2-m^2]
(k^2-\Lambda^2) n\cdot k}.
\end{eqnarray}

 The loop integration will become much easier to carry out if the large-$\Lambda$
limit can be taken before the integration. However, this operation
is only feasible if the integration is finite before and after
taking the large-$\Lambda$ limit. Therefore, we first successively
use the identity \cite{martin1}
\begin{eqnarray}
\frac{1}{(k+p)^2-m^2}=\frac{1}{k^2-m^2}-\frac{2k\cdot p+p^2}{(k^2-m^2)[(k+p)^2-m^2]}
\label{decom}
\end{eqnarray}
to reduce the superficial UV divergent degree of the integrand until
the  large-$\Lambda$ limit can be safely taken. For example, a term
in $\Sigma_{\rm NP}$ can be calculated as follows:
\begin{eqnarray}
&& \lim_{\Lambda\to\infty}\int\frac{d^dk}{(2\pi)^d}
\frac{\Lambda k_\mu}{(k^2-\Lambda^2)[(k+p)^2-m^2]}\nonumber\\
&=&\lim_{\Lambda\to\infty}\int\frac{d^dk}{(2\pi)^d}\frac{\Lambda k_\mu}{(k^2-\Lambda^2)}
\left[ \frac{1}{k^2-m^2}-\frac{2k\cdot p+p^2}{(k^2-m^2)[(k+p)^2-m^2]} \right]\nonumber\\
&=&-\lim_{\Lambda\to\infty}\int\frac{d^dk}{(2\pi)^d}\frac{\Lambda
k_\mu}{(k^2-\Lambda^2)} \frac{2k\cdot p+p^2}{(k^2-m^2)}
\left[\frac{1}{k^2-m^2}-\frac{2k\cdot p+p^2}{(k^2-m^2)[(k+p)^2-m^2]}
\right]\nonumber\\
&=&-\lim_{\Lambda\to\infty}\int\frac{d^dk}{(2\pi)^d}\left\{
\frac{2\Lambda k\cdot p k_\mu}{(k^2-\Lambda^2)(k^2-m^2)^2}
-\frac{\Lambda k_\mu (2 k\cdot p+p^2)^2}{(k^2-\Lambda^2)(k^2-m^2)^2} \right.\nonumber\\
&& \times \left.
\left[\frac{1}{k^2-m^2}-\frac{2k\cdot p+p^2}{(k^2-m^2)[(k+p)^2-m^2]} \right] \right\}\nonumber\\
&=&-\lim_{\Lambda\to\infty}\int\frac{d^3k}{(2\pi)^3}
\frac{2\Lambda k\cdot p k_\mu}{(k^2-\Lambda^2)(k^2-m^2)^2}\nonumber\\
&=&-\lim_{\Lambda\to\infty}\frac{2}{3}\Lambda p_\mu  \int\frac{d^3k}{(2\pi)^3}
\frac{k^2}{(k^2-\Lambda^2)(k^2-m^2)^2}\nonumber\\
&=&ip_\mu\,\lim_{\Lambda\to\infty}\left[-\frac{1}{12}\frac{\Lambda
(2\Lambda^3-3\Lambda^2m+m^3)} {(\Lambda^2-m^2)^2}
\right]=-\frac{i}{6\pi} p_\mu.
 \end{eqnarray}
In above calculation we have used the even and odd property of the
integrands. Other terms in $\Sigma_{\rm NP}$ can be evaluated in a
similar way. For the terms in $\Sigma_{\rm P}$ we first use the ML
prescription shown in (\ref{mlpres}) to deal with the spurious
light-cone pole.  Then we choose a convenient Lorentz frame for the
light-cone vector  $n_\mu$ to perform the loop integration in a
non-covariant way, and finally express the results in terms of a
Lorentz invariant functions with the light-cone vector $n_\mu$ and
its conjugate $n_\mu^\ast$. The explicit calculation techniques are
shown in Appendix \ref{appa}. Consequently, we obtain the fermionic
 self-energy at one-loop order,
\begin{eqnarray}
-i\Sigma^{(1)}(p,m,n) &=&\lim_{\Lambda\to\infty}\left\{\lim_{d\to
3}\left[-i\Sigma^{(1)}(p,m,n,\Lambda, d)\right]\right\}
\nonumber\\
&=&\frac{ie^2}{2\pi}\left[ \Lambda +\frac{2}{3}m -
\frac{5}{6}\left(p\hspace{-1.6mm}/ - m\right) +\frac{(n\cdot
p)n\hspace{-2.2mm}/^\ast -(n^\ast\cdot p)
n\hspace{-2.2mm}/}{n^\ast\cdot n} -\frac{m^2
n\hspace{-2.2mm}/}{n\cdot p}
\right.\nonumber\\
&&-\left. \frac{m\left(n\cdot p - m n\hspace{-2.2mm}/\right)}{n\cdot
p} \left(1-\frac{2(n^\ast\cdot p)(n\cdot p)}{m^2 (n^\ast\cdot n)}
\right)^{1/2} \right].
\label{selfe}
\end{eqnarray}

\section{Vertex Correction on Mass Shell at one-loop}
\label{osvertex}

In the following we consider the one-loop quantum correction for the
vertex $\psi-\overline{\psi}-A$ on the mass-shell of the fermion.
That is,
\begin{eqnarray}
&& -i\overline{u}(p^\prime)\Gamma^{(1)}_\mu (p^\prime,p, m, n)u(p) \nonumber\\
&=& \lim_{\Lambda\to\infty}\overline{u}(p^\prime)\left\{e^2\int
\frac{d^3k}{(2\pi)^3}\frac{\gamma_\rho
(k\hspace{-2mm}/+p\hspace{-1.8mm}/ ^\prime +m)\gamma_\mu
(k\hspace{-2mm}/+p\hspace{-1.8mm}/
+m)\gamma^\nu}{[(k+p^\prime)^2-m^2] [(k+p)^2-m^2]}
 \right.\nonumber\\
 &&\times \left. \frac{1}{k^2-\Lambda^2}\left[\frac{i\Lambda^2}{n\cdot k}\epsilon^{\nu\rho\lambda}
 n_\lambda-\Lambda g^{\nu\rho}
 +\frac{\Lambda}{n\cdot k}\left(k_\nu n_\rho+k_\rho n_\nu \right) \right]\right\} u(p)
 \nonumber\\
 &\equiv& -i\left( \Gamma_{[1]\mu}+\Gamma_{[2]\mu}+\Gamma_{[3]\mu}\right),
 \label{vertex1}
\end{eqnarray}
where the Dirac spinor $u(p)$ is a solution of the Dirac equation and $\overline{u}(p)$
is its conjugate,
\begin{eqnarray}
(p\hspace{-1.8mm}/-m)u(p)=0, ~~~~\overline{u}(p)(p\hspace{-1.8mm}/-m)=0.
\label{diraceq}
\end{eqnarray}
The three parts in (\ref{vertex1}) are listed as follows:
\begin{eqnarray}
-i\Gamma_{[1]\mu}&=&
\lim_{\Lambda\to\infty}\overline{u}(p^\prime)\left\{-\Lambda e^2\int
\frac{d^3k}{(2\pi)^3}\frac{\left[-k\hspace{-2mm}/ \gamma_\nu +2
(k+p^\prime)_{\nu}\right] \gamma_\mu \left[-\gamma_\nu
k\hspace{-2mm}/ + 2 (k+p)^\nu \right]}{(k^2-\Lambda^2)
[(k+p^\prime)^2-m^2] [(k+p)^2-m^2]}\right\}u(p); \nonumber\\
\\
-i\Gamma_{[2]\mu}&=&
\lim_{\Lambda\to\infty}\overline{u}(p^\prime)\left\{\Lambda e^2\int
\frac{d^3k}{(2\pi)^3} \frac{1}{(n\cdot k)
(k^2-\Lambda^2)[(k+p^\prime)^2-m^2]
[(k+p)^2-m^2]}\right. \nonumber\\
&&\times \left. \left[n\hspace{-2.4mm}/
(k\hspace{-2mm}/+p\hspace{-1.8mm}/ ^\prime +m)\gamma_\mu
(k\hspace{-2mm}/+p\hspace{-1.8mm}/ +m)k\hspace{-2mm}/ +
k\hspace{-2mm}/ (k\hspace{-2mm}/+p\hspace{-1.8mm}/ ^\prime
+m)\gamma_\mu (k\hspace{-2mm}/+p\hspace{-1.8mm}/
+m)n\hspace{-2.4mm}/\right] \right\}u(p)\nonumber\\
&=&\lim_{\Lambda\to\infty}\overline{u}(p^\prime)\left\{\Lambda
e^2\int \frac{d^3k}{(2\pi)^3} \frac{1}{(n\cdot k)
(k^2-\Lambda^2)[(k+p^\prime)^2-m^2]
[(k+p)^2-m^2]}\right. \nonumber\\
&&\times \left[\left(2 n\cdot (k+p^\prime)-k\hspace{-2mm}/
n\hspace{-2.4mm}/ \right) \gamma_\mu (k^2+2 k\cdot
p)\right.\nonumber\\
&& \left. \left. +(k^2+2 k\cdot p^\prime )\gamma_\mu \left(2 n\cdot
(k+p)-n\hspace{-2.4mm}/
k\hspace{-2mm}/ \right)\right]\right\}u(p);\\
-i\Gamma_{[3]\mu}&=&
\lim_{\Lambda\to\infty}\overline{u}(p^\prime)\left\{i\Lambda^2
e^2\int \frac{d^3k}{(2\pi)^3} \frac{1}{(n\cdot k) (k^2-\Lambda^2)
[(k+p^\prime)^2-m^2] [(k+p)^2-m^2]}\right. \nonumber\\
&&\times  \left. \epsilon_{\nu\rho\lambda}n^\lambda
\left[-k\hspace{-2.2mm}/ \gamma^\rho + 2 (k+ p^\prime )^\rho
\right]\gamma_\mu \left[-\gamma^\nu k\hspace{-2.2mm}/  + 2 (k+
p)^\nu \right] \right\} u(p).
\end{eqnarray}
In writing down $\Gamma_{[1]\mu}$, $\Gamma_{[2]\mu}$ and
$\Gamma_{[3]\mu}$, we have used the mass shell condition shown in
Eq.\,(\ref{diraceq}),
\begin{eqnarray}
\overline{u}(p^\prime)\gamma_\rho (k\hspace{-2mm}/+p\hspace{-1.8mm}/ ^\prime
+m) &=& \overline{u}(p^\prime) \left[-k\hspace{-2mm}/ \gamma_\rho
+ 2 (k+p^\prime )_\rho \right],\nonumber\\
(p\hspace{-1.8mm}/ + k\hspace{-2mm}/ +m )\gamma_\nu u (p) &=&
\left[- \gamma_\nu k\hspace{-2mm}/ + 2(k+p)_\nu \right] u(p).
\end{eqnarray}

In the following we calculate $\Gamma_{[1]\mu}$, $\Gamma_{[2]\mu}$
and $\Gamma_{[3]\mu}$.

\begin{itemize}
\item $\Gamma_{[1]\mu}$
\end{itemize}

$\Gamma_{[1]\mu}$ can be reduced to the following form with the
$\gamma$-matrix algebra (\ref{gamma}) and the mass shell condition
given in Eq.\,(\ref{diraceq}),
\begin{eqnarray}
&& -i\Gamma_{[1]\mu}=
\lim_{\Lambda\to\infty}\overline{u}(p^\prime)\left\{-\Lambda e^2
\int \frac{d^3k}{(2\pi)^3}
\frac{1}{(k^2-\Lambda^2)[(k+p^\prime)^2-m^2] [(k+p)^2-m^2]} \right.\nonumber\\
&&\times \left.\left[k^2\gamma_\mu -2k\hspace{-2.2mm}/ k_\mu +4
k\cdot (p^\prime +p)\gamma_\mu +4m k_\mu -4 k\hspace{-2mm}/
(p^\prime_\mu +p_\mu)+4p^\prime \cdot p\gamma_\mu
\right]\right\}u(p)
\end{eqnarray}
We can take the large-$\Lambda$ limit before evaluating the
integration for the term with the numerator $4p^\prime\cdot
p\gamma_\mu$, which vanishes after taking the large-$\Lambda$ limit.
As for other terms, we must first make the decomposition
(\ref{decom}) successively until it is feasible to take the
large-$\Lambda$ limit. It can be easily seen that the terms whose
numerator linear in $k_\mu$ vanishes:
\begin{eqnarray}
&& \lim_{\Lambda\to\infty}\int\frac{d^3k}{(2\pi)^3}
\frac{\Lambda k_\mu}{(k^2-\Lambda^2)
[(k+p^\prime)^2-m^2] [(k+p)^2-m^2]}\nonumber\\
&=& \lim_{\Lambda\to\infty}\int\frac{d^3k}{(2\pi)^3}
\frac{\Lambda k_\mu}{(k^2-\Lambda^2)(k^2-m^2)^2}
\left[1-\frac{2k\cdot p^\prime+p^{\prime 2}}{(k+p^\prime)^2-m^2} \right]
\left[1-\frac{2k\cdot p+p^2}{(k+p)^2-m^2} \right]\nonumber\\
&=&\lim_{\Lambda\to\infty}\int\frac{d^3k}{(2\pi)^3}
\frac{\Lambda k_\mu}{(k^2-\Lambda^2)(k^2-m^2)^2}
\left[-\frac{2k\cdot p^\prime+p^{\prime 2}}{(k+p^\prime)^2-m^2}
-\frac{2k\cdot p+p^2}{(k+p)^2-m^2}
\right.\nonumber\\
&&\left.
 +\frac{( 2k\cdot p^\prime+p^{\prime 2}) (2k\cdot p+p^2)}
 {[(k+p^\prime)^2-m^2 ][ (k+p^\prime)^2-m^2]}
\right]=0.
\end{eqnarray}

As for the first two terms whose numerators are quadratic in
$k_\mu$, we have from the decomposition (\ref{decom}),
\begin{eqnarray}
&& \lim_{\Lambda\to\infty}\int\frac{d^3k}{(2\pi)^3}\frac{\Lambda k_\mu k_\nu}{(k^2-\Lambda^2)
[(k+p^\prime)^2-m^2] [(k+p)^2-m^2]}\nonumber\\
&=& \lim_{\Lambda\to\infty}\int\frac{d^3k}{(2\pi)^3}\frac{\Lambda
k_\mu k_\nu}{(k^2-\Lambda^2)(k^2-m^2)^2} \left[1-\frac{2k\cdot
p^\prime }{(k+p^\prime)^2-m^2}-\frac{2k\cdot p}{(k+p)^2-m^2}
\right].
\end{eqnarray}
Hence only the first term survives after the large-$\Lambda$ limit.
Thus we obtain
\begin{eqnarray}
-i\Gamma_{[1]\mu}&=& \lim_{\Lambda\to\infty}\left[-\Lambda e^2 \int
\frac{d^3k}{(2\pi)^3}
\frac{\gamma_\mu k^2-2 k\hspace{-2mm}/ k_\mu}{(k^2-\Lambda^2)(k^2-m^2)^2} \right]\nonumber\\
&=& \lim_{\Lambda\to\infty}\left[-\frac{1}{3} \Lambda e^2\gamma_\mu
\int
\frac{d^3k}{(2\pi)^3}\frac{k^2}{(k^2-\Lambda^2)(k^2-m^2)^2} \right]\nonumber\\
&=&\lim_{\Lambda\to\infty}\left[-\frac{1}{3} e^2\gamma_\mu
\frac{i}{8\pi}\frac{\Lambda
(2\Lambda^3-3\Lambda^2m+m^3)}{(\Lambda^2-m^2)^2}\right]
=-\frac{ie^2}{12\pi}\gamma_\mu \label{gamma1mu}
\end{eqnarray}

\begin{itemize}
\item $\Gamma_{[2]\mu}$
\end{itemize}

To evaluate $\Gamma_{[2]\mu}$, we first separate it into the sectors
with and without the spurious light-cone singularity $(n\cdot
k)^{-1}$, and impose the mass shell conditions $p^{\prime
2}=p^2=m^2$. Then $\Gamma_{[2]\mu}$ takes the following form,
\begin{eqnarray}
-i\Gamma_{[2]\mu}&=&
\lim_{\Lambda\to\infty}\overline{u}(p^\prime)\left\{ \Lambda e^2
\int\frac{d^3k}{(2\pi)^3}\left[\frac{2\gamma_\mu}{(k^2-\Lambda^2)(k^2+2k\cdot
p^\prime )}
+\frac{2\gamma_\mu}{(k^2-\Lambda^2)(k^2+2k\cdot p)}\right.\right.\nonumber\\
&+& \frac{2n\cdot p^\prime \gamma_\mu}{n\cdot k (k^2-\Lambda^2)(k^2+2k\cdot p^\prime )}
+\frac{2n\cdot p\gamma_\mu}{n\cdot k (k^2-\Lambda^2)(k^2+2k\cdot p)}\nonumber\\
&-&\left. \left.\frac{k\hspace{-2.3mm}/ n\hspace{-2.4mm}/
\gamma_\mu}{n\cdot k (k^2-\Lambda^2)(k^2+2k\cdot p^\prime )}
-\frac{\gamma_\mu n\hspace{-2.4mm}/ k\hspace{-2.3mm}/ }{n\cdot k
(k^2-\Lambda^2)(k^2+2k\cdot p)}\right]\right\}u(p).
\end{eqnarray}
Note that the mass shell condition $p^{\prime 2}=p^2=m^2$ should not
be imposed on some terms until the integrations have been performed
in order to avoid the artificial infrared divergence caused by
implementing the mass shell condition. Using the formula listed in
Appendix \ref{appb}, we obtain
\begin{eqnarray}
-i\Gamma_{[2]\mu}
= \frac{1}{\pi}ie^2\gamma_\mu -
\frac{1}{2\pi}ie^2\frac{n\hspace{-2.4mm}/ ^\ast n_\mu-
n\hspace{-2.4mm}/ n_\mu^\ast}{n^\ast\cdot n}. \label{gamma2mu}
\end{eqnarray}

\begin{itemize}
\item $\Gamma_{[3]\mu}$
\end{itemize}

We first use the $\gamma$-matrix algebra,
$\epsilon_{\mu\nu\rho}\gamma^\rho = i/2\,[\gamma_\mu,\gamma_\nu]$,
to rewrite $\Gamma_{[3]\mu}$ and take the large-$\Lambda$ limit on
those feasible terms to simplify $\Gamma_{[3]\mu}$. Then there
appears
\begin{eqnarray}
-i\Gamma_{[3]\mu}&=&\lim_{\Lambda\to\infty}\overline{u}(p^\prime)\left[
\Lambda^2 e^2 \int\frac{d^3k}{(2\pi)^3}\frac{4k_\mu}{(k^2-\Lambda^2)
[(k+p^\prime)^2-m^2][(k+p)^2-m^2]}  \right]u(p)\nonumber\\
&+&\lim_{\Lambda\to\infty}\overline{u}(p^\prime)\left[ -\Lambda^2
e^2 \int\frac{d^3k}{(2\pi)^3}\frac{2k^2 n_\mu}{n\cdot k
(k^2-\Lambda^2)
[(k+p^\prime)^2-m^2][(k+p)^2-m^2]}\right]u(p)\nonumber\\
&-&\overline{u}(p^\prime)e^2\int\frac{d^3k}{(2\pi)^3}\frac{4 (p_\mu
k\hspace{-2mm}/ n\hspace{-2.2mm}/ +p_\mu^\prime n\hspace{-2.2mm}/
k\hspace{-2mm}/ ) -2m (k\hspace{-2mm}/ n\hspace{-2.2mm}/ \gamma_\mu
+\gamma_\mu n\hspace{-2.2mm}/ k\hspace{-2mm}/) -2 (n\cdot p
k\hspace{-2mm}/ \gamma_\mu+ n\cdot p^\prime \gamma_\mu
k\hspace{-2mm}/) }{n\cdot k (k^2+2k\cdot p^\prime)(k^2+2k\cdot
p)}u(p)\nonumber\\
&-&  \overline{u}(p^\prime)\left[4ie^2
\int\frac{d^3k}{(2\pi)^3}\frac{\epsilon^{\nu\rho\lambda} n_\lambda
(p_\nu k_\rho+ k_\nu p^{\prime}_{\rho}+p_\nu p^{\prime}_{\rho} )
\gamma_\mu} {n\cdot k (k^2+2k\cdot p^\prime)(k^2+2k\cdot p)}\right] u(p) \nonumber\\
&\equiv& -i\left[V_{(1)\mu }+V_{(2)\mu }+V_{(3)\mu }+V_{(4)\mu
}\right]\, . \label{gamma3mu}
\end{eqnarray}

Using the decomposition (\ref{decom}), taking the large-$\Lambda$
limit and then putting them on the mass shell, we can calculate
$V_{(1)\mu}$ and $V_{(2)\mu}$ as follows:
\begin{eqnarray}
-iV_{(1)\mu}&=& \lim_{\Lambda\to\infty}\overline{u}(p^\prime)
\left[\Lambda^2 e^2
\int\frac{d^3k}{(2\pi)^3}\frac{4k_\mu}{(k^2-\Lambda^2)
[(k+p^\prime)^2-m^2][(k+p)^2-m^2]}\right]\,u(p)\nonumber\\
&=& \overline{u}(p^\prime)\left\{-4e^2 \int\frac{d^3k}{(2\pi)^3}
\frac{k_\mu}{(k^2-m^2)^2}\left[ -\frac{2k\cdot p^\prime +p^{\prime
2}}{(k+p^\prime)^2-m^2}
\right.\right.\nonumber\\
&&\left.\left.-\frac{2k\cdot p +p^2}{(k+p)^2-m^2}+\frac{(2k\cdot
p^\prime +p^{\prime 2})(2k\cdot p +p^2)}
{[(k+p^\prime)^2-m^2][(k+p)^2-m^2]} \right]\right\}u(p)\nonumber\\
&=&
\overline{u}(p^\prime)\left[-4e^2\int\frac{d^3k}{(2\pi)^3}\frac{k_\mu}{(k^2+2k\cdot
p^\prime)
(k^2+2k\cdot p) }\right] u(p)\nonumber\\
&=&\overline{u}(p^\prime) \left[ \frac{ie^2}{4\pi}
\frac{p_\mu^\prime +p_\mu}{q}\ln\frac{1+q/(2m)}{1-q/(2m)}\right]
u(p). \label{vonemu}
\end{eqnarray}
\begin{eqnarray}
-i V_{(2)\mu}&=&
-\lim_{\Lambda\to\infty}\int\frac{d^3k}{(2\pi)^3}\left.\frac{2\Lambda^2
e^2n_\mu  k^2} {(k^2-\Lambda^2)(n\cdot k)
[(k+p^\prime)^2-m^2][(k+p)^2-m^2]} \right|_{p^2=p^{\prime 2}=m^2}\nonumber\\
&=&2e^2n_\mu \int\frac{d^3k}{(2\pi)^3}\frac{k^2}{(n\cdot
k)(k^2-m^2)^2}\left[
-\frac{2k\cdot p^\prime +p^{\prime 2}}{(k+p^\prime)^2-m^2}\right.\nonumber\\
&&\left.\left.-\frac{2k\cdot p +p^2}{(k+p)^2-m^2}
+\frac{(2k\cdot p^\prime +p^{\prime 2})(2k\cdot p +p^2)}
{[(k+p^\prime)^2-m^2][(k+p)^2-m^2]} \right]\right|_{p^2=p^{\prime 2}=m^2}\nonumber\\
&=& 2e^2n_\mu \int\frac{d^3k}{(2\pi)^3}\frac{k^2}{(n\cdot k)
(k^2+2k\cdot p^\prime)
(k^2+2k\cdot p) }= -2e^2n_\mu g^{\nu\rho}I_{\nu\rho}\nonumber\\
&=& -\frac{ie^2}{8\pi}n_\mu \left[ \frac{1}{n\cdot (p^\prime +p)}
\frac{4m^2-q^2}{q}\ln\frac{1+q/(2m)}{1-q/(2m)}
+m\left(\frac{1}{n\cdot p^\prime}+\frac{1}{n\cdot p} \right)\right.\nonumber\\
&&\left.+\frac{1}{n\cdot n^\ast}\left(\frac{n\cdot (p^\prime +p)}{m}
-2 \right) \frac{D(p^\prime)^{1/2}- D(p)^{1/2}}{n\cdot (p^\prime
-p)}\right]\, . \label{vtwomu}
\end{eqnarray}
In above equations, $q_\mu\equiv p_\mu^\prime-p_\mu$ and $D(p)\equiv
m^2 n\cdot n^\ast -2(n^\ast\cdot p)(n\cdot p)$. In addition, we have
used the integral formula (\ref{imunuform}) of $I_{\mu\nu}$ worked
out in Appendix \ref{appa}.

To show the explicit symmetry of $V_{(3)\mu}$ and $V_{(4)\mu}$ in
$p^\prime_\mu$ and $p_\mu$, we express
\begin{eqnarray}
p_\mu^\prime=\frac{1}{2}\left({\cal P}_\mu+q_\mu \right),~~
p_\mu=\frac{1}{2}\left({\cal P}_\mu+q_\mu \right)
\end{eqnarray}
in evaluating $V_{(3)\mu}$ and $V_{(4)\mu}$,  where ${\cal P}_\mu
\equiv 1/2 \left( p_\mu^\prime+p_\mu\right)$. Then
\begin{eqnarray}
-iV_{(3)\mu}&=&
-e^2\overline{u}(p^\prime)\left[4\left(p_\mu^\prime+p_\mu\right)
\int\frac{d^3k}{(2\pi)^3}\frac{1}{(k^2+2k\cdot
p^\prime)(k^2+2k\cdot p)} \right.\nonumber\\
&& +2q_\mu \int\frac{d^3k}{(2\pi)^3}\frac{n\hspace{-2.2mm}/
k\hspace{-2mm}/ - k\hspace{-2mm}/ n\hspace{-2.2mm}/ }{n\cdot
k(k^2+2k\cdot p^\prime) (k^2+2k\cdot p)} \nonumber\\
&& \left.-2m\int\frac{d^3k}{(2\pi)^3}\frac{k\hspace{-2mm}/
n\hspace{-2.2mm}/ \gamma_\mu+\gamma_\mu n\hspace{-2.2mm}/
k\hspace{-2mm}/ }{n\cdot k(k^2+2k\cdot p^\prime) (k^2+2k\cdot p)}
\right] u(p)
\end{eqnarray}
\begin{eqnarray}
-iV_{(4)\mu}&=& \overline{u}(p^\prime) \left[ -4ie^2\gamma_\mu
\epsilon^{\nu\rho\lambda}n_\lambda
q_\rho\int\frac{d^3k}{(2\pi)^3}\frac{k_\nu+1/2\,
(p_\nu^\prime+p_\nu)}{n\cdot k(k^2+2k\cdot p^\prime) (k^2+2k\cdot
p)} \right]u(p)
\end{eqnarray}

Using the integral formulas (\ref{threeint}) and (\ref{i2mutens})
for $I_2$ and $I_{2\mu}$, we have
\begin{eqnarray}
-iV_{(3)\mu}&=&-ie^2\left\{\frac{1}{2\pi} \left(p_\mu^\prime+ p_\mu
\right)\frac{1}{q}\ln\frac{1+q/(2m)}{1-q/(2m)} \right.\nonumber\\
&&+4 E_1\left[n\cdot q q_\mu-2 m^2 n_\mu-n\cdot (p^\prime+p)
m\gamma_\mu+m n\hspace{-2.2mm}/  \left(p_\mu^\prime+ p_\mu \right)
\right]\nonumber\\
&& +2 E_3 \left[q_\mu \left(n\hspace{-2.2mm}/
{n\hspace{-2.2mm}/}^\ast -{n\hspace{-2.2mm}/}^\ast
n\hspace{-2.2mm}/\right) +2m \left(n\hspace{-2.2mm}/ n^\ast_\mu
-{n\hspace{-2.2mm}/}^\ast n_\mu\right)-2m n\cdot n^\ast \gamma_\mu
\right]\, , \label{vthreemu}
\end{eqnarray}
\begin{eqnarray}
-iV_{(4)\mu}=-4 e^2\gamma_\mu\epsilon^{\nu\rho\lambda}n_\nu q_\rho
\left[\left(E_1+\frac{1}{2} I_2 \right) \left(p_\lambda^\prime+
p_\lambda \right)+E_3 n^\ast_\lambda\right]\, , \label{vfourmu}
\end{eqnarray}
where $I_2$, $E_1$ and $E_3$ are the Lorentz scalar functions
constructed from  $p_\mu$, $p_\mu^\prime$, $n_\mu$ and $n_\mu^\ast$
and are symmetric in $p_\mu$ and $p_\mu^\prime$, and their explicit
forms are given in (\ref{threeint}), (\ref{coefE1}) and
(\ref{coefE3}).

The one-loop quantum vertex function $\Gamma_\mu^{(1)}$ on mass
shell of the fermion in the light-cone gauge can be obtained by
summing up $\Gamma_{[1]\mu}$, $\Gamma_{[2]\mu}$ and
$\Gamma_{[3]\mu}$.

\section{Renormalization and Structure of Local Quantum Effective Action}
\label{renorma}

\subsection{Finite Renormalization of Gauge Field Propagator and Generation of Maxwell Term}

Eq.\,(\ref{vpt1}) shows that the vacuum polarization tensor
$\Pi_{\mu\nu}(p)$ is finite. The finite renormalization on the
propagator of the $U(1)$ CS gauge field can still be performed
according to the standard procedure. The inverse of CS gauge field
propagator up to one-loop level is
\begin{eqnarray}
\left[iG_{\mu\nu}^{(1)}(p) \right]^{-1} &=& \left[iG_{\mu\nu}^{(0)}(p)\right]^{-1}
-i\Pi_{\mu\nu}(p)\nonumber\\
&=& i\left[\epsilon_{\mu\nu\lambda}ip^\lambda \left(1-\Pi_{\rm
o}(p)\right) +\left( p^2 g_{\mu\nu}-p_\mu p_\nu \right)\Pi_{\rm
e}(p) +\frac{1}{\xi} n_\mu n_\nu \right].
  \end{eqnarray}
Hence
\begin{eqnarray}
 iG_{\mu\nu}^{(1)}&=& -i \frac{1-\Pi_{\rm o}(p)}{\left[1-\Pi_{\rm o}(p)\right]^2
 -p^2\left[\Pi_{\rm e}(p)\right]^2}
 \left[\frac{i}{p^2}\epsilon_{\mu\nu\rho}p^\rho
 -\frac{i}{(n\cdot p)p^2}(p_\mu \epsilon_{\nu\alpha\beta}-p_\nu
 \epsilon_{\mu\alpha\beta})p^\alpha n^\beta \right.\nonumber\\
 &&\left.-\frac{\Pi_{\rm e}(p)}{1-\Pi_{\rm o}(p)} g_{\mu\nu}
 +\frac{\Pi_{\rm e}(p)}{1-\Pi_{\rm o}(p)}\,
 \frac{1}{(n\cdot p)}(p_\mu n_\nu + p_\nu n_\mu)
  \right]\nonumber\\
  &=&  \frac{1}{n\cdot p}\epsilon_{\mu\nu\rho}n^\rho\,
    \frac{1-\Pi_{\rm o}(p)}{\left[1-\Pi_{\rm o}(p)\right]^2-p^2
    \left[\Pi_{\rm e}(p)\right]^2}\nonumber\\
   && +i\left[ g_{\mu\nu}-\frac{1}{n\cdot p}(p_\mu n_\nu + p_\nu n_\mu)\right]
   \frac{\Pi_{\rm e}(p)}{\left[1-\Pi_{\rm o}(p)\right]^2
   -p^2\left[\Pi_{\rm e}(p)\right]^2}\nonumber\\
 &=&  \frac{1}{n\cdot p} \epsilon_{\mu\nu\rho}n^\rho
\frac{1}{1+\Pi_1 (p)} +i\left[ g_{\mu\nu}-\frac{1}{n\cdot
p}\left(p_\mu n_\nu + p_\nu n_\mu\right)\right]\Pi_2 (p),
  \label{fprop}
\end{eqnarray}
where the Martin identity (\ref{mid1}) is employed and
\begin{eqnarray}
\Pi_1 (p)&=& -\Pi_{\rm o} (p)-\frac{p^2\Pi_{\rm e}^2 (p)}{1-\Pi_{\rm o} (p)},\nonumber\\
\Pi_2 (p)&=& \frac{\Pi_{\rm e}(p)}{[1-\Pi_{\rm o}(p)]^2-p^2\Pi_{\rm e}^2(p)}.
\end{eqnarray}

We choose the renormalization condition that at $p=0$
 \begin{eqnarray}
 \Pi_{1{\rm R}}(0)=0,
 \end{eqnarray}
 and define the wave function renormalization constant of
the CS gauge field in the usual way,
\begin{eqnarray}
Z_3^{-1}=1+\Pi_1(0)=1-\Pi_{\rm o}(0).
\end{eqnarray}
This gives
\begin{eqnarray}
Z_3=1+\frac{e^2}{4\pi}.
\end{eqnarray}
Consequently, the one-loop renormalized propagator of the $U(1)$ CS
gauge field (i.e., up to the order $e^2$) is
\begin{eqnarray}
iG_{\mu\nu R}^{\rm (1)}(p)&=& Z_3^{-1}\left[iG_{\mu\nu}^{\rm (1)}(p)\right]\nonumber\\
&=& \frac{1}{n\cdot p} \epsilon_{\mu\nu\rho}n^\rho
\frac{1}{1+\Pi_{1R} (p)}
+i\left[ g_{\mu\nu}-\frac{1}{n\cdot p}\left(p_\mu n_\nu + p_\nu n_\mu\right)\right]\Pi_{2R} (p),
\end{eqnarray}
where at one-loop level,
\begin{eqnarray}
\Pi_{1R} (p) &=& \Pi_{1} (p)- \Pi_{1} (0)=\Pi_{\rm o}(0)-\Pi_{\rm o}(p)\nonumber\\
&=&\frac{e^2}{4\pi}\left[ 1-\frac{m}{p}\ln \frac{1+p/(2m)}{1-p/(2m)}\right]; \nonumber\\
\Pi_{2R} (p) &=& \Pi_{\rm e} (p)=\frac{e^2}{4\pi}\frac{1}{m} \left[
\frac{m^2}{p^2}-\frac{m}{p}\left(\frac{1}{4}+ \frac{m^2}{p^2}\right)
\ln \frac{1+p/(2m)}{1-p/(2m)}\right]. \label{ptformfactor}
\end{eqnarray}
Eq.\,(\ref{ptformfactor}) shows
\begin{eqnarray}
\Pi_{2R}(0)=-\frac{e^2}{4\pi}\frac{1}{3m}\neq 0.
\end{eqnarray}
This fact means that the parity-even Maxwell term in the CS spinor
electrodynamics is generated
 by quantum correction, which is a general feature of the CS gauge theory
 coupled with fermions \cite{kogan}.

\subsection{Renormalization of Fermionic Propagator}

Eq.\,(\ref{selfe}) shows that the self-energy is composed of the
light-cone vector dependent part $\Sigma^{(1)}_{\rm (I)} $ and the
independent one $\Sigma^{(1)}_{\rm (D)}$:
\begin{eqnarray}
 \Sigma^{(1)}(p,m,n,\Lambda)&=& \Sigma^{(1)}_{\rm (I)}(p,m,\Lambda)
+\Sigma^{(1)}_{\rm (D)} (p,m,n), \nonumber\\
 \Sigma^{(1)}_{\rm
(I)}(p,m,\Lambda) &=& \frac{e^2}{2\pi}\left[ -\Lambda -\frac{2}{3}m
+ \frac{5}{6}\left(p\hspace{-1.6mm}/ -m\right)\right],
\label{selfeid}\\
\Sigma^{(1)}_{\rm (D)} (p,m,n)&=& -\frac{e^2}{2\pi}
\left\{-\frac{(n\cdot p)n\hspace{-2.2mm}/^\ast -(n^\ast\cdot p)
n\hspace{-2.2mm}/}{n^\ast\cdot n}+\frac{m^2
n\hspace{-2.2mm}/}{n\cdot p}
\right.\nonumber\\
&&\left. +\frac{1}{2}\frac{m}{n\cdot p}\left(1-\frac{2(n^\ast\cdot
p)(n\cdot p)}{m^2 (n^\ast\cdot n)} \right)^{1/2} \left[
n\hspace{-2.3mm}/ \left( p\hspace{-1.6mm}/ -m\right)+\left(
p\hspace{-1.6mm}/ -m\right)n\hspace{-2.3mm}/ \right] \right\}.
\label{selfede}
\end{eqnarray}

We impose the following mass-shell renormalization condition on the
light-cone vector independent part $\Sigma_{\rm (I)\,R}(p)$:
\begin{eqnarray}
\left.\Sigma_{\rm (I)\,R}(p)\right|_{p\hspace{-1.3mm}/ =m_{\rm
R}}=0, ~~~ \left.\frac{\partial}{\partial p\hspace{-1.8mm}/ }
\Sigma_{\rm (I)\,R} (p) \right|_{p\hspace{-1.3mm}/ =m_{\rm R}}=0.
\end{eqnarray}
Then $\Sigma_{\rm (I)}(p,m,\Lambda)$ has the following expansion
around $p\hspace{-1.7mm}/ =m_{\rm R}$,
\begin{eqnarray}
\Sigma_{\rm (I)}(p,m,\Lambda) = \delta m -
\left(Z_2^{-1}-1\right)\left(p\hspace{-1.7mm}/-m_{\rm R}\right)+
Z_2^{-1}\Sigma_{\rm (I)\,R}(p), \label{renorform}
\end{eqnarray}
where  $Z_2$ is the wave function constant of the fermion.


Eqs.\,(\ref{selfeid}) and (\ref{renorform}) yield that
 the renormalized fermionic mass, the wave function renormalization
 constant of the fermion
 and the light-cone vector independent part of one-loop fermionic self-energy are as follows,
\begin{eqnarray}
m_R&=& m-\delta m=\frac{e^2}{2\pi}\left(\Lambda +\frac{2}{3}m \right),\nonumber\\
Z_2 &=& 1+\frac{e^2}{4\pi}\,\frac{5}{3}\, ,
 \nonumber\\
\Sigma_{\rm (I)\,R} &=& 0.
\end{eqnarray}

The light-vector dependent sector $\Sigma_{\rm (D)}(p,m,n)$ is
finite. We shall show that combined with the light-cone vector
dependent sector in the vertex correction, it contributes to a gauge
invariant quantum effective action specific to the light-cone gauge.

\subsection{Finitely Renormalized On-shell Vertex Correction
and Arising of Anomalous Magnetic Moment of Fermion}

Collecting the results shown in  Eqs.\,(\ref{vertex1}),
(\ref{gamma1mu}), (\ref{gamma2mu}), (\ref{gamma3mu}),
(\ref{vonemu}), (\ref{vtwomu}), (\ref{vthreemu}) and
(\ref{vfourmu}), we see that the the  on-shell vertex correction at
one-loop is finite, and consists of the light-cone vector
independent sector $\Gamma^{(1)}_{{\rm (I)}\mu}$ and the dependent
sector $\Gamma_{{\rm (D)}\mu}^{(1)}$:
\begin{eqnarray}
\Gamma^{(1)}_{{\rm (I)}\mu}&=&\frac{e^2}{4\pi}\left\{
\left[-\frac{11}{3}+\frac{2m}{q}\ln\frac{1+q/(2m)}{1-q/2m}\right]\gamma_\mu
-\frac{1}{q}\ln\frac{1+q/(2m)}{1-q/2m}\,i\epsilon_{\mu\nu\rho}q^\nu\gamma^\rho\right\},
\label{indepver}\\
\Gamma_{{\rm (D)}\mu}^{(1)}&=& \frac{e^2}{2\pi}
\frac{n\hspace{-2.4mm}/ ^\ast n_\mu- n\hspace{-2.4mm}/
n_\mu^\ast}{n^\ast\cdot n}+\mbox{non-polynomial terms in $p_\mu$ and
$p_\mu^\prime$ }. \label{depver}
\end{eqnarray}
In writing down Eq.\,(\ref{indepver}), we have used the
three-dimensional analogue of the Gordon identity,
\begin{eqnarray}
\overline{u}(p^\prime)
\left(p_{\mu}^\prime+p_\mu\right)u(p)=\overline{u}(p^\prime)\left(2m\gamma_\mu
-i\epsilon_{\mu\nu\rho}q^\nu\gamma^\rho\right)u(p).
\end{eqnarray}

To perform the finite renormalization on the vertex correction, we
choose the renormalized light-cone vector independent sector
$\Gamma^{(1)R}_{{\rm (I)}\mu}$ to satisfy
\begin{eqnarray}
\left.\Gamma^{(1)R}_{{\rm (I)}\mu}(p^\prime,p)\right|_{p^{\prime
2}=p^2=m^2, q_\mu=0}=0\, , \label{vdefcon}
\end{eqnarray}
and define the vertex renormalization constant $Z_1$ as follows,
\begin{eqnarray}
\Gamma^{(1)}_{{\rm
(I)}\mu}(p^\prime,p)=\left(Z_1^{-1}-1\right)\gamma_\mu+Z_1^{-1}\Gamma^{(1)R}_{{\rm
(I)}\mu}(p^\prime,p)\, . \label{renormvertex}
\end{eqnarray}
Then  from Eqs.\,(\ref{indepver}), (\ref{vdefcon}) and
(\ref{renormvertex}) we obtain the vertex renormalization constant
at one-loop level:
\begin{eqnarray}
Z_1^{-1}-1 &=& -\frac{e^2}{4\pi}\frac{5}{3}, \nonumber\\
Z_1 &=& 1+\frac{e^2}{4\pi}\frac{5}{3}
\end{eqnarray}
It is equal to $Z_2$, the wave function renormalization constant  of
the fermion, which is a direct consequence of the Ward identity
(\ref{verserela1}) or (\ref{verserela3}).

 According to Eq.\,(\ref{renormvertex}), the light-cone vector
independent radiative corrections of the vertex at one-loop is
\begin{eqnarray}
\Gamma^{(1)R}_{{\rm (I)}\mu}(p^\prime,p)=
-\gamma_\mu+Z_1\left[\Gamma^{(1)}_{{\rm (I)}\mu}(p^\prime,p)
+\gamma_\mu\right]=\gamma_\mu
F_1(q^2)+i\epsilon_{\mu\nu\rho}q^\nu\gamma_\rho F_2(q^2)
\end{eqnarray}
where
\begin{eqnarray}
F_1(q^2)&=&\frac{e^2}{4\pi}\left[-2+\frac{2m}{q}\ln\frac{1+q/(2m)}{1-q/(2m)}
 \right]\, , \nonumber\\
 F_2(q^2)&=& -\frac{1}{q}\ln\frac{1+q/(2m)}{1-q/(2m)}.
 \label{vertexff}
\end{eqnarray}
Eq.\,(\ref{vertexff}) shows that  at the renormalization point
$q^2=0$, the form factor $F_2(q^2)$ does not vanish,
\begin{eqnarray}
F_2 (0)=-\frac{1}{m}\, .
\end{eqnarray}
This actually gives rise to the analogue of Schwinger's result for
the anomalous magnetic moment of the fermion in the CS spinor
electrodynamics. The term with tensor structure
$\epsilon_{\mu\nu\rho}q^\nu \gamma_\rho$ and the form factor
$F_2(q^2)$ leads to an
 interaction Hamiltonian at a higher order when the fermions are in a
slowly varying $U(1)$ CS gauge field (since $q_\mu\rightarrow 0$),
\begin{eqnarray}
\Delta {\cal
H}=-\frac{e^2}{4\pi}\frac{1}{m}\epsilon_{\mu\nu\rho}\overline{\psi}(x)\gamma_\rho
\overline{\psi}(x)\partial^\nu A^\mu (x)
=\frac{e^2}{8\pi}\frac{1}{m}\overline{\psi}(x)\sigma_{\mu\nu}\overline{\psi}(x)F^{\mu\nu}(x)\,.
\end{eqnarray}
This result coincides with that obtained in the covariant gauge
\cite{kogan}.

\subsection{Contribution to Local Quantum Effective Action
from Light-cone Vector Dependent Terms}

We now turn to the light-cone vector dependent terms appearing in
the fermionic self-energy and in the on-shell vertex correction.
Eqs.\,(\ref{selfede}) and (\ref{depver}) lead to the following
light-cone vector dependent local fermionic quantum effective action
 at one-loop order,
\begin{eqnarray}
\Gamma_{\rm (D)}^{(1)}&=&\frac{e^2}{2\pi}\frac{1}{n\cdot
n^\ast}\left[ i\overline{\psi} \left(n\hspace{-2.4mm}/ ^\ast n_\mu
\partial^\mu-n\hspace{-2.4mm}/ n^\ast_\mu
\partial^\mu\right)\psi- \overline{\psi}
\left(n\hspace{-2.4mm}/ ^\ast n_\mu A^\mu -n\hspace{-2.4mm}/
n^\ast_\mu A^\mu\right) \psi  \right]\nonumber\\
&=&\frac{e^2}{2\pi}\frac{1}{n\cdot n^\ast}i\overline{\psi} \left(
n\hspace{-2.4mm}/ ^\ast n_\mu D^\mu - n\hspace{-2.4mm}/ n^\ast_\mu
D^\mu\right)\psi\, ,
\end{eqnarray}
where $D_\mu=\partial_\mu-ie A_\mu$ is the covariant derivative.
$\Gamma_{\rm (D)}^{(1)}$ is invariant under the $U(1)$ gauge
transformation listed in Eq.\,(\ref{uonegt}). It should be
emphasized that this is precisely analogous to the result of a
four-dimensional non-Abelian gauge theory coupled with fermions  in
the light-cone gauge \cite{banaso}

The non-polynomial terms  in the external momenta given in
Eqs.\,(\ref{selfede}) and (\ref{depver}) will contribute to the
non-local sector of the light-cone vector dependent quantum
effective action for the fermion. Unfortunately, unlike the pure
non-Abelian CS gauge theory in the light-cone gauge, which has no
dimensional parameter \cite{martin2,glbook}, we are unable to
extract out the explicit form of the non-local light-cone vector
dependent quantum effective action due to the complications of those
non-polynomial terms.


\section{Summary and Conclusion}
\label{summary}

 A complete investigation in the perturbation theory of  Chern-Simons spinor
 electrodynamics in the light-cone gauge ($n\cdot A=0$, $n^2=0$) at
 one-loop order has been made. We have calculated
 the vacuum polarization tensor, fermionic self-energy and on-shell vertex
  correction, and further performed the one-loop
renormalization to define the quantum theory. The peculiar features
of quantum corrections of Chern-Simons spinor electrodynamics in the
light-cone gauge have been revealed. Two typical quantum effects in
CS spinor electrodynamics, the generation of the parity-even Maxwell
term and the arising of anomalous magnetic moment of the fermion
from quantum corrections, have been reproduced as in the case of the
covariant gauge fixing. We have also shown that as a consequence of
the Ward identities in the light-cone gauge, the wave function
renormalization constant of the fermion is equal to the vertex
renormalization constant. Further, we have displayed the structure
of local  quantum effective action for the fermion, and found that
its light-cone vector dependent sector is explicit gauge invariant.
Especially, it takes exactly the same form as that in a
four-dimensional gauge theory coupled with fermions in the
light-cone gauge. This result is a natural consequence of the Ward
identities for the CS spinor electrodynamics in the light-cone
gauge. Therefore, the covariance of S-matrix elements will be
achieved.

The result summarized above has not only verified the applicability
of the ML-prescription to three-dimensional gauge theory in the
presence of fermions, but also shown the gauge independence of
Chern-Simons type of gauge theory in evaluating gauge invariant
physical observables.

\acknowledgments

\noindent A partial support from the start-up research grant of
Nipissing University is acknowledged. This work is in memory of the
late Professor George Leibbrandt, who led the author into the field
of non-covariant gauge theory.

\appendix

\section{Feynman Integral with Spurious Light-Cone Gauge Singularity
in Leibbrandt-Mandelstam Prescription} \label{appa}

In this appendix we show how the Feynman integrals containing the
spurious light-cone pole in three dimensions are evaluated with the
ML prescription. Actually,
 only the following five types of integrals containing the pole
  are needed for evaluating
 the fermionic self-energy and on-shell vertex correction:
 \begin{eqnarray}
 iI_1 &=& \int\frac{d^3k}{(2\pi)^3}\frac{1}{n\cdot k [(k+p)^2-m^2]};\nonumber\\
 iI_2 &=& \int \frac{d^3k}{(2\pi)^3}
 \frac{1}{n\cdot k (k^2+2k\cdot p^\prime) (k^2+2k\cdot p)}; \nonumber\\
 I_{1\mu}&=& \lim_{d\to 3}\int  \frac{d^dk}{(2\pi)^d}
 \frac{k_\mu}{(n\cdot k) (k^2+2k\cdot p)}; \nonumber\\
 I_{\mu\nu} &=&\lim_{d\to 3}\int  \frac{d^dk}{(2\pi)^d}
 \frac{k_\mu k_\nu}{n\cdot k (k^2+2k\cdot p^\prime) (k^2+2k\cdot p)};\nonumber\\
  I_{2\mu} &=&\lim_{d\to 3}\int  \frac{d^dk}{(2\pi)^d}
  \frac{k_\mu }{n\cdot k (k^2+2k\cdot p^\prime) (k^2+2k\cdot p)}.
 \end{eqnarray}
  We adopt the procedure illustrated
in Ref.\,\cite{banaso} rather than the exponential parametrization
used in three-dimensional non-covariant gauge theory
\cite{glbook,martin2,martin3}. For the convenience  of calculation,
we choose the Lorentz frame such that
\begin{eqnarray}
n=\left(n_0,0,n_2 \right), ~~~n^\ast=\left(n_0,0,-n_2\right), ~~~n_0>0.
\end{eqnarray}
The superficially covariant three-vector notation will be restored
at the end of calculation. Since the light-gauge vectors $n_\mu$ and
$n_\mu^\ast$ satisfy $n^2=n^{\ast 2}=0$, there exist
\begin{eqnarray}
n_2&=&\pm n_0, ~~ \kappa \equiv \frac{n_2}{n_0}=\pm 1, ~~ n_0^2=n_2^2=
\frac{1}{2}n\cdot n^\ast.\nonumber\\
 p^2_0-p_2^2 &=&  (p_0+\kappa p_2) (p_0-\kappa p_2)\nonumber\\
&=& \frac{1}{n_0^2}\left( n_0p_0+n_2p_2\right)
\left(n_0p_0-n_2p_2\right)=\frac{2(n^\ast\cdot p) (n\cdot p) }{n^\ast\cdot n}.
\end{eqnarray}

\begin{itemize}

\item \textbf{Evaluation of $I_1$}

\begin{eqnarray}
iI_1 &=& \int\frac{d^3k}{(2\pi)^3}
\frac{n^\ast\cdot k}{[(n\cdot k)(n^\ast\cdot k)+i\epsilon] [(k+p)^2-m^2]}\nonumber\\
&\equiv&\frac{1}{n_0}\int\frac{d^3k}{(2\pi)^3}
\frac{k_0+\kappa k_2}{(k_0^2-k_2^2)[(k+p)^2-m^2]}\nonumber\\
&=& \frac{1}{n_0}\int_0^1 dx \frac{1}{(2\pi)^3}
\int_{-\infty}^{\infty}dk_1 \int_{-\infty}^{\infty}dk_2
\int_{-\infty}^{\infty}dk_0 \,\left[\left(k_0+\kappa k_2 \right)\right.\nonumber\\
&& \left.\times \frac{1}
 {[(k_0+p_0x)^2-(k_2+p_2x)^2-(k_1+p_1)^2x+(p_0^2-p_2^2)x(1-x)-m^2x]^2}\right]\nonumber\\
 &=& -\frac{i}{8\pi}\frac{p_0+\kappa p_2}{n_0}
 \frac{1}{m}\int_0^1dx \frac{1}{\left[ 1-(p^2_0-p_2^2)(1-x)/m^2\right]^{1/2}}\nonumber\\
 &=&-\frac{i}{4\pi}\frac{m}{n_0(p_0-\kappa p_2)}
 \left[1-\left(1-\frac{p_0^2-p_2^2}{m^2} \right)^{1/2} \right]\nonumber\\
 &=&-\frac{i}{4\pi}\frac{m}{n\cdot p}
 \left[1-\left(1-\frac{2(n^\ast\cdot p)(n\cdot p)} {m^2(n^\ast\cdot n)} \right)^{1/2} \right].
\end{eqnarray}

\item \textbf{Calculation of $I_2$}

\begin{eqnarray}
iI_2 &=& \int \frac{d^3k}{(2\pi)^3}
\frac{1}{n\cdot k (k^2+2k\cdot p^\prime) (k^2+2k\cdot p)}\nonumber\\
&\equiv& \int \frac{d^3k}{(2\pi)^3}
\frac{n^\ast\cdot k}{\left[(n^\ast\cdot k) (n\cdot k)+i\epsilon\right]
(k^2+2k\cdot p^\prime) (k^2+2k\cdot p)}
\nonumber\\
&=&\frac{1}{2n_0}\frac{1}{(2\pi)^3}\int_{-\infty}^{\infty}dk_1\int_{-\infty}^{\infty}dk_2
\int_{-\infty}^{\infty}dk_0\int_0^1dx \int_0^1 dy 2y  (k_0+\kappa
k_2)
\nonumber\\
&& \times \left[ \frac{1}{\left\{[(k^2+2k\cdot
p^\prime)x+(k^2+2k\cdot p)(1-x)]y +(k_0^2-k_2^2) (1-y)
\right\}^3}\right.\nonumber\\
&+&\left.  \frac{1}{\left\{[(k^2+2k\cdot p)x+(k^2+2k\cdot p^\prime
)(1-x)]y +(k_0^2-k_2^2) (1-y) \right\}^3}
\right]\nonumber\\
&=&\frac{i}{16\pi}\int_0^1dx \frac{n\cdot (p^\prime +p)}{n\cdot (p+q
x)\,n\cdot (p^\prime - q x)}
\frac{1}{[m^2-q^2 x(1-x)]^{1/2}}\nonumber\\
&& -\frac{i}{16\pi}\int_0^1 dx \left[\frac{1}{n\cdot (p+q
x)}\right.\nonumber\\
&&\times \frac{1}{\left[m^2-q^2 x(1-x)-2n^\ast\cdot (p+ q
x)\,n\cdot (p+q x)/( n^\ast\cdot n) \right]^{1/2}}\nonumber\\
&&+\frac{1}{n\cdot (p^\prime-q x)}\nonumber\\
&&\times\left. \frac{1}{\left[m^2-q^2 x(1-x)-2n^\ast\cdot (
p^\prime-q x)\,n\cdot (p^\prime-q x)/( n^\ast\cdot n)
\right]^{1/2}}\right],
  \label{threeint}
\end{eqnarray}
which shows that $I_2$ is symmetric in  $p_\mu$ and $p_\mu^\prime$.

\item \textbf{Calculating $I_{1\mu}$}

According to the Lorentz covariance, $I_{1\mu}$ has the following
tensor structure,
\begin{eqnarray}
 I_{1\mu}&=& \lim_{d\to 3}\int  \frac{d^dk}{(2\pi)^d}
\frac{k_\mu}{(n\cdot k) (k^2+2k\cdot p)}=iK_1\,p_{\mu}+ iK_2\,n_\mu
K_2+iK_3\,n^{\ast}_\mu ,
 \label{i1muten}
\end{eqnarray}
where the undetermined coefficients $K_1$, $K_2$ and $K_3$ are the
functions of Lorentz scalars constructed from $p_\mu$, $n_\mu$ and
$n_\mu^\ast$. Then making the projections of $I_{1\mu}$ on $p_\mu$,
$n_\mu$ and $n_\mu^\ast$, respectively, we have
\begin{eqnarray}
X &\equiv & I_{1\mu}p^\mu=iK_1\, m^2 +iK_2\, n\cdot p +iK_3\,n^\ast\cdot p \nonumber\\
&=& \lim_{d\to 3}\int  \frac{d^dk}{(2\pi)^d} \frac{k\cdot p}{(n\cdot k) (k^2+2k\cdot p)};
\label{i1mux}\\
Y &\equiv & I_{1\mu}n^\mu= iK_1\, n\cdot p +iK_3\,n\cdot n^\ast \nonumber\\
&=& \lim_{d\to 3}\int  \frac{d^dk}{(2\pi)^d} \frac{1}{k^2+2k\cdot p}; \label{i1muy}\\
Z &\equiv & I_{1\mu}n^{\star\mu}=i K_1\,n^\ast\cdot p +iK_2\,n\cdot n^\ast \nonumber\\
&=& \lim_{d\to 3}\int  \frac{d^dk}{(2\pi)^d} \frac{n^\ast\cdot
k}{(n\cdot k) (k^2+2k\cdot p)}. \label{i1muz}
\end{eqnarray}
It is straightforward to evaluate $X$, $Y$ and $Z$ using the ML
prescription and taking into account the mass shell condition
$p^2=m^2$. Note that in the regularized $d$-dimensions,
$k_\mu=(k_0,k_{\bot},k_2)$ and $k_{\bot}$ has $d-2$ components. Then
\begin{eqnarray}
&& Y = \lim_{d\to 3}\int  \frac{d^dk}{(2\pi)^d} \frac{1}{k^2+2k\cdot p}
=\lim_{d\to 3}\int  \frac{d^dk}{(2\pi)^d} \frac{1}{k^2-m^2}=\frac{i}{4\pi} m.\\
&& X =\lim_{d\to 3}\int
\frac{d^dk}{(2\pi)^d} \frac{k\cdot p}{(n\cdot k)(k^2+2k\cdot p)}\nonumber\\
&=&\lim_{d\to 3}\int  \frac{d^dk}{(2\pi)^d}
\frac{(n^\ast\cdot k)k\cdot p}
{[(n^\ast\cdot k)(n\cdot k)+i\epsilon ](k^2+2k\cdot p)}\nonumber\\
&=&\frac{1}{n_0}\lim_{d\to 3}\int_0^1dx\frac{1}{(2\pi)^d}
\int_{-\infty}^{\infty} dk_0
\int_{-\infty}^{\infty} dk_2\int d^{d-2}k_{\bot}
\left[(k_0+\kappa k_2)\right.\nonumber\\
&\times& \left. \frac{k_0p_0-k_2p_2-k_{\bot}p_{\bot}}
{\left[(k_0+p_0 x)^2-(k_2+p_2x)^2-( k_{\bot}+p_{\bot})^2x+ (p_0^2-p_2^2)x(1-x)-m^2x \right]^2}\right]
\nonumber\\
&=& \frac{i}{4\pi}\frac{m (p_0+\kappa p_2)}{n_0}\int_0^1dx
\left[1-\frac{p_0^2-p_2^2}{m^2} (1-x) \right]^{1/2}\nonumber\\
&=& \frac{i}{6\pi}\frac{1}{n\cdot p}\left[ m^3-\frac{D(p)^{3/2}}{(n^\ast\cdot n)^{3/2}}\right].\\
&& Z =\lim_{d\to 3}\int  \frac{d^dk}{(2\pi)^d}
\frac{n^\ast\cdot k}{(n\cdot k)(k^2+2k\cdot p)}\nonumber\\
&=& \lim_{d\to 3}\int  \frac{d^dk}{(2\pi)^d}
\frac{(n^\ast\cdot k)^2}{[(n^\ast\cdot k)(n\cdot k)+i\epsilon](k^2+2k\cdot p)} \nonumber\\
&=& \frac{i}{8\pi}\frac{(p_0+\kappa p_2)^2}{m}\int_0^1dx
\frac{x}{\left[1-(1-x)(p_0^2-p_2^2)/m^2 \right]^{1/2}}\nonumber\\
&=&\frac{i}{4\pi} m \left[\frac{n^\ast\cdot p}{n\cdot p}
-\frac{1}{3}\frac{m^2 n^\ast\cdot n}{(n\cdot p)^2}
+\frac{1}{3}\frac{D(p)^{3/2}}{m (n\cdot p)^2(n^\ast\cdot n)^{1/2}}\right],
\end{eqnarray}
where $D(p)=m^2n^\ast\cdot n-2 (n^\ast\cdot p)(n\cdot p)$. Solving
the system of algebraic equations for $K_1$, $K_2$ and $K_3$ listed
in (\ref{i1mux}), (\ref{i1muy}) and (\ref{i1muz}), we have
\begin{eqnarray}
K_1 &=&\frac{1}{4\pi}\frac{1}{D(p)}
\left( n^\ast\cdot n X-n^\ast\cdot p Y- n\cdot pZ\right)\nonumber\\
&=&\frac{1}{4\pi}\frac{m}{n\cdot p}\left[1-\frac{D(p)^{1/2}}{m(n\cdot n^\ast)^{1/2}} \right].\\
K_2 &=& \frac{Z}{n^\ast\cdot n}-\frac{n^\ast\cdot p}{n\cdot n^\ast}K_1\nonumber\\
&=&\frac{1}{12\pi}\left(\frac{m}{n\cdot p}\right)^2\left[-m
+\frac{D(p)^{1/2}}{(n^\ast\cdot n)^{1/2}}
\left(1+\frac{(n^\ast\cdot p)(n\cdot p)}{m^2n\cdot n^\ast}\right) \right].\\
K_3 &=& \frac{Y}{n\cdot n^\ast}-\frac{n^\ast\cdot p}{n\cdot
n^\ast}K_1=\frac{1}{4\pi}\frac{D(p)^{1/2}}{(n\cdot n^\ast)^{3/2}}.
\label{i1mucoe}
\end{eqnarray}

\item \textbf{Evaluating $I_{\mu\nu}$}

$I_{\mu\nu}$ is invariant under the exchanges $\mu\leftrightarrow\nu
$ and $p_\mu\leftrightarrow p^\prime_\mu$, respectively. Therefore,
the general tensor structure of $I_{\mu\nu}$ should be the following
form:
 \begin{eqnarray}
 I_{\mu\nu} &=&\lim_{d\to 3}\int  \frac{d^dk}{(2\pi)^d}
 \frac{k_\mu k_\nu}{n\cdot k (k^2+2k\cdot p^\prime) (k^2+2k\cdot p)}\nonumber\\
 &=& iC_1 (p_\mu^\prime p_\nu^\prime +p_\mu p_\nu)
 +iC_2  (p_\mu^\prime p_\nu+p_\mu p_\nu^\prime)+iC_3
 \left[n_\mu^\ast (p_\nu^\prime+p_\nu)+ n_\nu^\ast (p_\mu^\prime+p_\mu)\right]\nonumber\\
 && +iC_4 \left[n_\mu (p_\nu^\prime+p_\nu)+ n_\nu (p_\mu^\prime+p_\mu)\right]
 +iC_5n_\mu^\ast n_\nu^\ast +iC_6 n_\mu n_\nu \nonumber\\
 &&+iC_7 \left(n_\mu^\ast n_\nu+n_\nu^\star n_\mu \right) +iC_8g_{\mu\nu},
 \label{imunuform}
 \end{eqnarray}
 where $C_i$ ($i=1,2,\cdots,8$)  are functions of the Lorentz
 scalars constructed from $p_\mu$, $p_\mu^\prime$, $n_\mu$ and $n_\mu^\ast$,
and are symmetric in $p^\prime_\mu$ and  $p_\mu$. Then contracting
$I_{\mu\nu}$ with the vector $n^\nu$, and using Eq.\,(\ref{kmuns}),
we obtain
 \begin{eqnarray}
 C_1&=&C_2, ~~ C_8=-C_4 n\cdot (p^\prime +p)-C_7 n\cdot n^\prime, ~~
 C_5=-\frac{n\cdot (p^\prime +p)}{n\cdot n^\ast}C_3,\nonumber\\
 C_1&=& -\frac{1}{n\cdot (p^\prime +p)}
 \left[\frac{1}{16\pi}\frac{1}{q}\ln \frac{1+q/(2m)}{1-q/(2m)} +C_3 (n\cdot n^\ast) \right]
\end{eqnarray}
Further, $I_{\mu\nu}n^\mu n^\nu$ and Eq.\,(\ref{kmuns}) determine
that $C_3=0$. Hence
\begin{eqnarray}
C_1&=& C_2= -\frac{1}{n\cdot (p^\prime
+p)}\left[\frac{1}{16\pi}\frac{1}{q}
\ln \frac{1+q/(2m)}{1-q/(2m)}\right];\nonumber\\
C_5&=&0.
\label{vc1}
\end{eqnarray}
Consequently, $I_{\mu\nu}$ becomes
\begin{eqnarray}
 I_{\mu\nu} &=&\lim_{d\to 3} \int  \frac{d^dk}{(2\pi)^d}
 \frac{k_\mu k_\nu}{n\cdot k (k^2+2k\cdot p^\prime) (k^2+2k\cdot p)}\nonumber\\
 &=& iC_1(p_\mu^\prime+p_\mu)(p_\nu^\prime+p_\nu)+iC_4\left[ n_\mu (p_\nu^\prime+p_\nu)+
  n_\nu (p_\mu^\prime+p_\mu)\right]\nonumber\\
  && +iC_6 n_\mu n_\nu+ C_7 \left(n_\mu^\ast n_\nu+n_\nu^\ast n_\mu \right)
  -\left[iC_4 n\cdot (p^\prime+p)+iC_7 n\cdot n^\ast\right]g_{\mu\nu}.
  \label{imunu}
\end{eqnarray}
To evaluate  the scalar coefficients $C_1$, $C_4$, $C_6$ and $C_7$, we consider
$I_{\mu\nu} (p^{\prime\nu}-p^\nu)$,
\begin{eqnarray}
&& I_{\mu\nu} (p^{\prime\nu}-p^\nu)= p_\mu^\prime\left(-2iC_4 n\cdot
p-iC_7n\cdot n^\ast\right)
+p_\mu\left(2iC_4 n\cdot p^\prime+iC_7n\cdot n^\ast\right)\nonumber\\
 && +n_\mu \left[ iC_6 n\cdot (p^\prime-p)+ iC_7n^\ast\cdot (p^\prime-p) \right]
 +n^\ast_\mu iC_7  n\cdot (p^\prime-p)\nonumber\\
 &=&
 \lim_{d\to 3} \int  \frac{d^dk}{(2\pi)^d}
 \frac{k_\mu k\cdot (p^{\prime}-p)}{n\cdot k (k^2+2k\cdot p^\prime) (k^2+2k\cdot p)}\nonumber\\
&=&\frac{1}{2}\left[\lim_{d\to 3}
\int  \frac{d^dk}{(2\pi)^d}\frac{k_\mu}{n\cdot k (k^2+2k\cdot p)}
-\lim_{d\to 3} \int  \frac{d^dk}{(2\pi)^d}\frac{k_\mu}{n\cdot k (k^2+2k\cdot p^\prime)} \right].
\end{eqnarray}
Using the results (\ref{i1muten}) and (\ref{i1mucoe}) of $I_{1\mu}$,
we obtain the following algebraic equations:
\begin{eqnarray}
2C_4 n\cdot p^\prime+C_7 n\cdot n^\ast &=& \frac{1}{8\pi}\frac{m}{n\cdot p}
\left[ 1-\frac{D(p)^{1/2}}{m(n\cdot n^\ast )^{1/2}} \right],\\
2C_4 n\cdot p+C_7 n\cdot n^\ast &=& \frac{1}{8\pi}\frac{m}{n\cdot p^\prime}
\left[ 1-\frac{D(p^\prime)^{1/2}}{m(n\cdot n^\ast)^{1/2}} \right],\\
C_7 n\cdot (p^\prime-p)&=& -\frac{1}{8\pi}\frac{D(p^\prime)^{1/2}
-D(p)^{1/2}}{(n\cdot n^\ast)^{1/2}},\\
 C_6 n\cdot (p^\prime-p)+ C_7n^\ast\cdot (p^\prime-p) &=& K_2(p)-K_2(p^\prime),
\end{eqnarray}
which yield
\begin{eqnarray}
C_4 &=& \frac{1}{16\pi}\frac{m}{(n\cdot p)(n\cdot p^\prime)} +
\frac{i}{16\pi}\frac{1}{(n\cdot n^\ast)^{1/2}}
\frac{D(p^\prime)^{1/2}-D(p)^{1/2}}{n\cdot (p^\prime-p)};\nonumber\\
C_6 &=& -\frac{1}{24\pi}\frac{m^3}{(n\cdot p^\prime)(n\cdot p)}
\left(\frac{1}{n\cdot p^\prime}+ \frac{1}{n\cdot p}\right)\nonumber\\
&-& \frac{1}{24\pi}\frac{1}{(n^\ast\cdot n)^{1/2}}\frac{m^2}{n\cdot
(p^\prime-p)} \left[\frac{D(p^\prime)^{1/2}}{(n\cdot p^\prime)^2}
-\frac{D(p)^{1/2}}{(n\cdot p)^2} \right]\nonumber\\
&-&\frac{1}{24\pi}\frac{1}{(n^\ast\cdot n)^{3/2}}\frac{1}{n\cdot
(p^\prime-p)} \left[\frac{n^\ast\cdot p^\prime}{n\cdot
p^\prime}D(p^\prime)^{1/2}
-\frac{n^\ast\cdot p}{n\cdot p}D(p)^{1/2} \right]\nonumber\\
&+& \frac{1}{8\pi}\frac{1}{(n^\ast\cdot n)^{3/2}} \frac{n^\ast \cdot
(p^\prime-p) }{[n\cdot (p^\prime-p)]^2}
\left[D(p^\prime)^{1/2}-D(p)^{1/2}\right];\nonumber\\
C_7 &=& -\frac{1}{8\pi}\frac{1}{(n\cdot n^\ast )^{3/2}}
\frac{D(p^\prime)^{1/2}-D(p)^{1/2}}{n\cdot (p^\prime-p)}.\label{vcs}
\end{eqnarray}
Then $I_{\mu\nu}$ is given by Eqs.\,(\ref{vc1}), (\ref{imunu}) and (\ref{vcs}).

\item \textbf{Calculation of $I_{2\mu}$}

We calculate $I_{2\mu}$ in a similar way as evaluating $I_{1\mu}$,
whose tensor structure takes the following form,
\begin{eqnarray}
 I_{2\mu}&=& \int  \frac{d^3k}{(2\pi)^3}
 \frac{k_\mu}{(n\cdot k) (k^2+2k\cdot p)(k^2+2k\cdot p^\prime)}\nonumber\\
 &=& iE_1\,(p^\prime_{\mu}+p_\mu)+i E_2\,n_\mu +i E_3\,n^{\ast}_\mu .
 \label{i2mutens}
\end{eqnarray}
where $E_i$, $i=1,2,3$ are functions of the Lorentz scalars
constructed from $p_\mu$, $p_\mu^\prime$, $n_\mu$ and $n_\mu^\ast$,
and are symmetric in $p_\mu$ and $p^\prime_\mu$. Projecting
$I_{2\mu}$ on
 $n^\mu$, $(p^{\prime\mu}-p^\mu)$, and $(p^{\prime\mu}+p^\mu)$, respectively, and using
 the mass-shell condition, $p^2=p^{\prime 2}=m^2$,
 we have
\begin{eqnarray}
U &\equiv& n^\mu I_{2\mu}=E_1 n\cdot (p^{\prime}+p)+n\cdot n^\ast E_3\nonumber\\
&=& \int  \frac{d^3k}{(2\pi)^3} \frac{1}{(k^2+2k\cdot p)(k^2+2k\cdot p^\prime)}\, ,\label{equu}\\
V &\equiv &  (p^{\prime\mu}-p^\mu )I_{2\mu}
=E_2 n\cdot (p^{\prime}-p)+E_3 n^\ast\cdot (p^{\prime}-p)\nonumber\\
&=&\int  \frac{d^3k}{(2\pi)^3}
\frac{k\cdot (p^{\prime}-p)}{(n\cdot k) (k^2+2k\cdot p)(k^2+2k\cdot p^\prime)}\nonumber\\
&=&\frac{1}{2}\int  \frac{d^3k}{(2\pi)^3}
\frac{1}{(n\cdot k) (k^2+2k\cdot p)}-\frac{1}{2}\int
\frac{d^3k}{(2\pi)^3} \frac{1}{(n\cdot k) (k^2+2k\cdot p^\prime)}\, ,\label{equv}\\
W &=& (p^{\prime\mu}+p^\mu )I_{2\mu}
=E_1 (4m^2-q^2)+E_2 n\cdot (p^{\prime}+p)+E_3 n^\ast\cdot (p^{\prime}+p)\nonumber\\
&=&\int  \frac{d^3k}{(2\pi)^3} \frac{k\cdot (p^{\prime}+p)}{(n\cdot
k) (k^2+2k\cdot p)(k^2+2k\cdot p^\prime)}\, . \label{equw}
\end{eqnarray}
The scalar function $U$ can be calculated straightforwardly and is
given in Eq.\,(\ref{twop}). Further, the scalar function $V$ is
obtained from $I_1$ as follows,
\begin{eqnarray}
V= \frac{1}{2} \left[I_1(p)-I_1 (p^\prime)\right]\, .
 \end{eqnarray}
Finally, the scalar function $W$ can be evaluated from $I_1$ and
$g^{\mu\nu}I_{\mu\nu}$ by the following algebraic operations,
 \begin{eqnarray}
 && \int  \frac{d^3k}{(2\pi)^3}
 \frac{k\cdot (p^{\prime}+p)}{(n\cdot k) (k^2+2k\cdot p)(k^2+2k\cdot p^\prime)}
 =\frac{1}{2}\int\frac{d^3k}{(2\pi)^3} \frac{1}{(n\cdot k) (k^2+2k\cdot p)}\nonumber\\
 &+&\frac{1}{2}\int\frac{d^3k}{(2\pi)^3} \frac{1}{(n\cdot k) (k^2+2k\cdot p^\prime)}
 -\int\frac{d^3k}{(2\pi)^3} \frac{k^2}{(n\cdot k) (k^2+2k\cdot p)(k^2+2k\cdot p^\prime)}\nonumber\\
 &=& \frac{1}{2}I_{1}(p)+\frac{1}{2}I_{1}(p^\prime)-g^{\mu\nu}I_{\mu\nu}.
 \end{eqnarray}
Thus  $E_1$, $E_2$ and $E_3$ can be determined by solving the system
of algebraic equations (\ref{equu})-(\ref{equw}),
\begin{eqnarray}
E_1 &=&\frac{1}{N}
\left\{2\left(n\cdot p^\prime n^\ast\cdot p-n\cdot p n^\ast\cdot p^\prime \right)U\right.\nonumber\\
&& \left.-n^\ast\cdot n
\left[n\cdot (p^\prime +p)V-n\cdot (p^\prime -p)W \right]\right\}\, , \label{coefE1} \\
E_2 &=& \frac{1}{N}
\left\{\left[n\cdot(p^\prime +p)\,n^\star\cdot (p^\prime +p)-(4m^2-q^2)
n^\ast\cdot n \right]V\right.\nonumber\\
&&\left.+n^\ast\cdot (p^\prime -p)
\left[(4m^2-q^2)U- n\cdot(p^\prime +p)W\right]\right\} \, ,\label{coefE2}\\
E_3 &=& \frac{1}{N} \left\{-n\cdot (p^\prime -p)(4m^2-q^2)U
-\left[n\cdot(p^\prime +p)\right]^2V\right.\nonumber\\
&&\left.+n\cdot(p^\prime +p)\,n\cdot (p^\prime -p)W \right\},\label{coefE3}
\end{eqnarray}
where the denominator $N$ reads
\begin{eqnarray}
N=2n\cdot(p^\prime +p)\left(n\cdot p^\prime n^\ast\cdot p-n\cdot p n^\ast\cdot p^\prime \right)
-n\cdot n^\ast n\cdot(p^\prime -p)(4m^2-q^2).
\label{denomN}
 \end{eqnarray}
$I_{2\mu}$ is thus fixed from Eqs.(\ref{i2mutens})-(\ref{denomN}).

\end{itemize}

\section{Integration Formula } \label{appb}

We list in this appendix the integration formulas needed for
evaluating the vacuum polarization tensor, fermionic self-energy and
on-shell vertex correction. In the following, $q_\mu =
p_\mu^\prime-p_\mu$, $q=|q|$, $n=(n_\mu)=(n_0, {\bf n})$, and
$n^{\ast}=\left(n^{\ast}_\mu\right)=(n_0,-{\bf n})$.
\begin{eqnarray}
&& \int\frac{d^3k}{(2\pi)^3}\frac{1}{(k^2-m^2)[(k+p)^2-m^2]}
=\frac{i}{8\pi}\frac{1}{p}\ln\left[\frac{1+p/(2m)}{1-p/(2m)}\right].\\
&& \lim_{d\to 3}\int\frac{d^dk}{(2\pi)^d}\frac{k_\mu}{(k^2-m^2)[(k+p)^2-m^2]}
=-\frac{i}{16\pi}\frac{p_\mu}{p}\ln\frac{1+p/(2m)}{1-p/(2m)}.\\
&& \lim_{d\to 3}\int\frac{d^dk}{(2\pi)^d}\frac{k_\mu k_\nu}{(k^2-m^2)[(k+p)^2-m^2]}
\nonumber\\
&=& \frac{i}{16\pi}m \left\{\left[1+\frac{m}{p}
\left(1-\frac{p^2}{m^2} \right)\ln\frac{1+p/(2m)}{1-p/(2m)} \right]g_{\mu\nu}\right.\nonumber\\
&&+\left. \left[1+\frac{m}{p}\left(\frac{3}{4}
\frac{p^2}{m^2}-1 \right)\ln\frac{1+p/(2m)}{1-p/(2m)} \right]\frac{p_\mu p_\nu}{p^2}\right\}.\\
&& \lim_{\Lambda\to\infty}\int\frac{d^3k}{(2\pi)^3}\frac{\Lambda k_\mu}
{(k^2-\Lambda^2)[(k+p)^2-m^2]}\nonumber\\
&=&\lim_{\Lambda\to\infty}\left[-\int\frac{d^3k}{(2\pi)^3}
\frac{2\Lambda k\cdot p k_\mu}{(k^2-\Lambda^2) (k^2-m^2)^2}\right]\nonumber\\
&=& \lim_{\Lambda\to\infty}\left[-\frac{i}{12}p_\mu
\frac{\Lambda (2\Lambda^3-3\Lambda^2m+m^3)}{(\Lambda^2-m^2)^2} \right]=-\frac{i}{6\pi} p_\mu.\\
&& \lim_{\Lambda\to\infty}\int\frac{d^3k}{(2\pi)^3}\frac{\Lambda}
{(k^2-\Lambda^2)[(k+p)^2-m^2]}= \lim_{\Lambda\to\infty}\int\frac{d^3k}{(2\pi)^3}\frac{\Lambda}
{(k^2-\Lambda^2)(k^2-m^2)}\nonumber\\
&=&\lim_{\Lambda\to\infty}\int\frac{d^3k}{(2\pi)^3}\frac{\Lambda}
{(k^2-\Lambda^2)(k^2-m^2)}=\frac{i}{4\pi}.\\
&& \lim_{\Lambda\to\infty}\int\frac{d^3k}{(2\pi)^3}\frac{\Lambda^2}
{(k^2-\Lambda^2)[(k+p)^2-m^2]}
=\lim_{\Lambda\to\infty}\int\frac{d^3k}{(2\pi)^3}\frac{\Lambda^2}
{(k^2-\Lambda^2)(k^2-m^2)}\nonumber\\
&&+p^2\int\frac{d^3k}{(2\pi)^3}\frac{1}{(k^2-m^2)^2}
-\int\frac{d^3k}{(2\pi)^3}\frac{(2k\cdot
p+p^2)^2}{(k^2-m^2)^2[(k+p)^2-m^2]}\nonumber\\
&& =\frac{i}{4\pi}\Lambda.\\
&& \left.\int\frac{d^3k}{(2\pi)^3}\frac{1}{(k^2+2k\cdot p^\prime)
(k^2+2k\cdot p)}\right|_{p^{\prime 2}=p^2=m^2} =
\frac{i}{8\pi}\frac{1}{q}\ln\frac{1+q/(2m)}{1-q/(2m)}.\label{twop}\\
&& \left.\int\frac{d^3k}{(2\pi)^3}\frac{k_\mu}{(k^2+2k\cdot p^\prime)
(k^2+2k\cdot p)}\right|_{p^{\prime 2}=p^2=m^2}\nonumber\\
&=& -\frac{i}{16\pi}\frac{p_\mu^\prime+p_\mu}{q}\ln\frac{1+q/(2m)}{1-q/(2m)}.
\label{kmuns}\\
&& \lim_{\Lambda\to\infty}\int\frac{d^3k}{(2\pi)^3}\frac{\Lambda k_\mu}
{n\cdot k (k^2-\Lambda^2)[(k+p)^2-m^2]}
=\frac{i}{4\pi}\frac{n_\mu^\ast}{n\cdot n^\ast}.\\
&& \lim_{\Lambda\to\infty}\int\frac{d^3k}{(2\pi)^3}\frac{\Lambda k^2}
{n\cdot k (k^2-\Lambda^2)[(k+p)^2-m^2]}\nonumber\\
&=& \lim_{\Lambda\to\infty}\left[-\int\frac{d^3k}{(2\pi)^3}
\frac{2\Lambda k^2 k\cdot p} {n\cdot k (k^2-\Lambda^2) (k^2-m^2)^2}\right]
=-\frac{i}{2\pi}\frac{n^\ast\cdot p}{n\cdot n^\ast}.\\
&& \int\frac{d^3k}{(2\pi)^3}\frac{1}{(n\cdot k) (k^2+2k\cdot p)}=
-\frac{i}{4\pi}\frac{m}{n\cdot p}
\left[1-\left(1-\frac{2(n\cdot p)(n^\ast \cdot p) }{m^2(n\cdot n^\ast)} \right)^{1/2} \right].
\label{lightver3}
\end{eqnarray}

\section{Ward Identities in the Light-Cone Gauge} \label{appc}

The generating functional of the CS spinor electrodynamics in the
light-cone gauge is
 \begin{eqnarray}
 Z[J,\eta,\overline{\eta}]=\frac{1}{\cal N}\int {\cal D}\overline{\psi}{\cal D}\psi
 {\cal D}A \exp \left[i\int d^3x \left({\cal L}+\overline{\eta}\psi+\overline{\psi}\eta+
 J_\mu A^\mu \right) \right],
\end{eqnarray}
where the Lagrangian density ${\cal L}$ is given in
Eq.\,(\ref{sedeq1}), and $\overline{\eta}$, $\eta$ and $J_\mu$ are
the auxiliary external sources for $\psi$, $\overline{\psi}$ and
$A_\mu$, respectively. Note that $\overline{\eta}$ and $\eta$ are
the Grassmann variables. $Z[J,\eta,\overline{\eta}]$ is invariant
under the following gauge transformation,
 \begin{eqnarray}
 \psi^\prime (x)=e^{ie\theta (x)}\psi (x), ~~~\overline{\psi}^\prime (x)
 =\overline{\psi}e^{-ie\theta (x)},~~~A_\mu^\prime (x)=A_\mu (x)+\partial_\mu \theta (x).
 \label{uonegt}
 \end{eqnarray}
That is,
\begin{eqnarray}
\delta Z &=& \frac{1}{\cal N}\int {\cal D}\overline{\psi}{\cal D}\psi
 {\cal D}A\left\{ \exp \left[i\int d^3x \left({\cal L}+\overline{\eta}\psi+\overline{\psi}\eta+
 J_\mu A^\mu \right) \right]\right.\nonumber\\
 &&\times \left. i\int d^3y\left(-\frac{1}{\xi}n^\nu A_\nu n^\mu \partial_\mu\theta
 -ie\theta \overline{\eta}\psi +ie \theta\overline{\psi}\eta+
 J^\mu \partial^\mu \theta\right)\right\}=0.
\end{eqnarray}
Replacing $A_\mu (x)$, $\psi (x)$ and $\overline{\psi}(x)$ by the
functional derivatives $\delta Z/\delta J^\mu (x)$, $\delta Z/\delta \overline{\eta} (x)$
and  $\delta Z/\delta {\eta} (x)$, respectively, we obtain the identity,
\begin{eqnarray}
\left[\frac{1}{\xi}n^\lambda n^\mu\partial_\lambda \frac{\delta}{i\delta J^\mu (x)}
-ie\overline{\eta}(x)\frac{\delta}{i\delta \overline{\eta}(x)}+ie\eta (x)
\frac{\delta}{i\delta \overline{\eta}(x)}-\partial_\nu J^\nu (x) \right]Z=0.
\label{zwardi}
\end{eqnarray}
The corresponding Ward identity for the generating functional $W\equiv -i\ln Z$ of the
connected Green functions can be straightforwardly derived due to the linearity of the
the functional derivative operator in (\ref{zwardi}),
\begin{eqnarray}
\left[\frac{1}{\xi}n^\lambda n^\mu\partial_\lambda \frac{\delta}{i\delta J^\mu (x)}
-ie\overline{\eta}(x)\frac{\delta}{i\delta \overline{\eta}(x)}+ie\eta (x)
\frac{\delta}{i\delta \overline{\eta}(x)}-\partial_\mu J^\mu (x) \right]W=0.
\label{wwardi}
\end{eqnarray}
Acting $\delta/[i\delta J_\nu (y)]$ on the identity (\ref{wwardi})
and then setting the external sources $J_\mu$, $\eta$ and
$\overline{\eta}$ equal to zero, we obtain the Ward identity for the
two-point function of gauge field,
\begin{eqnarray}
&& \left.\left[\frac{1}{\xi}n^\lambda n^\mu \partial^x_\lambda
\frac{\delta^2}{i\delta J^\mu (x)i\delta J^\nu (y)}+i\partial_\mu^x
\delta^{(3)}(x-y) \right]W\right|_{J^\mu=\eta=\overline{\eta}=0}=0,\nonumber\\
&& n^\lambda n^\mu \partial^x_\lambda\left[iG_{\mu\nu}(x-y) \right]
=-i{\xi}\partial_\nu^x\delta^{(3)}(x-y).
\label{wigp}
\end{eqnarray}
In momentum space it reads as
\begin{eqnarray}
n^\mu G_{\mu\nu}(p)=-i{\xi}\frac{p_\nu}{n\cdot p}. \label{wardidgp}
\end{eqnarray}
Eq.\,(\ref{wardidgp}) implies that the tensor structure of two-point
function of the $U(1)$ CS gauge field is
\begin{eqnarray}
iG_{\mu\nu}(p)=A(p^2,n\cdot p)\epsilon_{\mu\nu\rho}n^\rho
+B(p, n\cdot p)\left[ g_{\mu\nu}-\frac{1}{n\cdot p}
\left(p_\mu n_\nu+p_\nu n_\mu \right)\right]+\xi\frac{p_\mu p_\nu}{(n\cdot p)^2}.
\end{eqnarray}

Further, acting $\delta/i\delta \overline{\eta}(y)$ and
$\delta/i\delta {\eta}(z)$ on the identity successively, and then
letting all the external sources equal to zero, we can obtain the
Ward identity relating the three-point function $\langle A_\mu (x)
\psi (y)\overline{\psi}(z) \rangle_{\rm C}$ and two-point function
$\langle \psi(x)\overline{\psi}(y)\rangle$:
\begin{eqnarray}
 && \left[\frac{1}{\xi}n^\mu n^\nu\partial_\nu^x
 \frac{\delta^3}{i\delta J^\mu (x)i\delta\overline{\eta}(y)
 i\delta\eta (z)}- e\delta^{(3)}(x-y)
 \frac{\delta^2}{i\delta\overline{\eta}(x)i\delta\eta (z)}\right.\nonumber\\
 &&\left.\left. + e\delta^{(3)}(x-z)\frac{\delta^2}{i\delta\overline{\eta}(y)i\delta\eta (x)}
 \right]W\right|_{J^\mu=\eta=\overline{\eta}=0}=0,\nonumber\\
 && \frac{1}{\xi}n^\mu n^\nu\partial_\nu^x
 \left\langle A_\mu (x)\psi (y)\overline{\psi}(z) \right\rangle_{\rm C}
 -e\delta^{(3)}(x-y)\left\langle \psi (x)\overline{\psi}(z)  \right\rangle\nonumber\\
&& +e\delta^{(3)}(x-z)\left\langle \psi (y)\overline{\psi}(x)  \right\rangle=0,
\label{verself}
\end{eqnarray}
where the subscript $C$ denotes the connected part of the
three-point function $\left\langle A_\mu (x)\psi
(y)\overline{\psi}(z) \right\rangle$. We further make
one-particle-irreducible decomposition on the connected three-point
function $\langle A_\mu (x)\psi(y)\overline{\psi}(z) \rangle_{\rm
C}$, and then Eq.\,(\ref{verself}) becomes
\begin{eqnarray}
&& \frac{1}{\xi}n^\mu n^\nu\partial_\nu^x \int d^3x^\prime d^3 y^\prime d^3 z^\prime
\left[iG_{\mu\lambda}(x-x^\prime)\right]\left[iS(y-y^\prime)\right]\left[iS(z-z^\prime)\right]
\Gamma_\lambda (x^\prime, y^\prime, z^\prime)\nonumber\\
&& -e\delta^{(3)}(x-y)\left[iS(x-z)\right]
+e\delta^{(3)}(x-z)\left[iS(y-x)\right]=0
\end{eqnarray}

Inserting (\ref{wigp}) and cutting-off the external legs, we obtain
the identity between the gauge field-fermion-fermion vertex function
and two-point function of the fermion,
\begin{eqnarray}
i\partial^x_\mu \Gamma^\mu (x,y,z)=\left[ iS(z-x)\right]^{-1}\delta^{(3)}(x-y)-
\left[ iS(x-y)\right]^{-1}\delta^{(3)}(x-z),
\end{eqnarray}
which is identical to the case in covariant gauge. In momentum space
it reads
\begin{eqnarray}
 q^\mu\Gamma_\mu \left[p^\prime,p, -(p^\prime+p)\right]
 = S^{-1}(p^\prime)-S^{-1}(p),
 \label{vsidentity}
\end{eqnarray}
where $q_\mu  \equiv  p^\prime_\mu -p_\mu$. Further, using the fact
that the perturbative quantum correction is the quantum fluctuation
around a classical background,
\begin{eqnarray}
\Gamma_\mu (p^\prime,
p) &=& \gamma_\mu+\Lambda_\mu(p^\prime,p),\nonumber\\
S^{-1}(p)&=& p\hspace{-2.2mm}/-m-\Sigma(p),
\end{eqnarray}
we finally obtain the identity relating the vertex correction and
fermionic self-energy.
\begin{eqnarray}
q^\mu \Lambda_\mu (p^\prime, p)=\left(p^{\prime\mu}-p^\mu
\right)\Lambda_\mu (p^\prime, p)=-\left[\Sigma (p^\prime)-\Sigma (p)
\right] \label{verserela1}
\end{eqnarray}
It is equivalent to
\begin{eqnarray}
\Lambda_\mu (p)=\lim_{p^{\prime}_\mu\rightarrow p_\mu}\Lambda_\mu
(p^\prime, p)=-\frac{\partial}{\partial p^\mu} \Sigma (p),
\label{verserela2}
\end{eqnarray}
which implies
\begin{eqnarray}
\Gamma_\mu (p)= -\frac{\partial}{\partial p^\mu} S^{-1} (p)
\label{verserela3}
\end{eqnarray}
The identity (\ref{verserela2}) or (\ref{verserela3})  leads to
$Z_1=Z_2$ as in the case of a covariant gauge.

\end{document}